\documentclass{article_saj}
\usepackage{graphicx,saj,multicol,subeqnarray}
\usepackage{natbib}
\usepackage{float}
\usepackage{xcolor}
\usepackage{widetext}
\usepackage{url}
\usepackage{bm}
\usepackage{tikz} 
\usepackage{pifont} 
\usepackage{amsfonts}
\usepackage{amssymb}
\usepackage{amsmath,upgreek}
\usepackage{titlesec}
\titlelabel{\thetitle.\quad}
\definecolor{xlinkcolor}{cmyk}{1,0.6,0,0}
\usepackage[bookmarks=false,         
     pdfnewwindow=true,      
     colorlinks=true,    
     linkcolor=xlinkcolor,     
     citecolor=xlinkcolor,     
     filecolor=xlinkcolor,  
     urlcolor=xlinkcolor,      
final=true
]{hyperref}

\def\papertype{\ \hfill\ Editorial}

\def\udc{52}
\begin{document}
\parindent=.5cm
\baselineskip=3.8truemm
\columnsep=.5truecm
\newenvironment{lefteqnarray}{\arraycolsep=0pt\begin{eqnarray}}
{\end{eqnarray}\protect\aftergroup\ignorespaces}
\newenvironment{lefteqnarray*}{\arraycolsep=0pt\begin{eqnarray*}}
{\end{eqnarray*}\protect\aftergroup\ignorespaces}
\newenvironment{leftsubeqnarray}{\arraycolsep=0pt\begin{subeqnarray}}
{\end{subeqnarray}\protect\aftergroup\ignorespaces}
%


\markboth{\eightrm A review on Stellar CMEs} 
{\eightrm M. Leitzinger \& P. Odert {\lowercase{\eightit{}}}}

\begin{strip}

{\ }

\vskip-1cm

\publ

\type

{\ }


\title{\uppercase{Stellar Coronal Mass Ejections}}


\authors{M. Leitzinger$^{1,\dagger}$ and P. Odert$^{1,\dagger}$}

\vskip3mm


\address{$^{1}$Institute of Physics/IGAM, University of Graz, Universit\"atsplatz 5, 8010 Graz, Austria}


\Email{martin.leitzinger@uni-graz.at; petra.odert@uni-graz.at}




\dates{May 18, 2020}{June 1, 2020}


\summary{Stellar coronal mass ejections (CMEs) are a growing research field, especially during the past decade. The large number of so far detected exoplanets raises the open question for the CME activity of stars, as CMEs may strongly affect exoplanetary atmospheres. In addition, as CMEs contribute to stellar mass- and angular momentum loss and are therefore relevant for stellar evolution, there is need for a better characterization of this phenomenon. In this article we review the different methodologies used up to now to attempt the detection of stellar CMEs. We discuss the limitations of the different methodologies and conclude with possible future perspectives of this research field.}


\end{strip}

\tenrm


\section{\uppercase{Introduction}}
\label{intro}
\indent

\footnotetext[2]{  Authors in alphabetical order, as M. Leitzinger and P. Odert contributed equally to this manuscript.}

Solar and stellar magnetic activity manifests itself in its most energetic forms in so-called flares, being characterized in a light curve as a rapid increase of intensity followed by a longer decay, and so-called coronal mass ejections (CMEs), being characterized as an expulsion of (partly) ionized magnetized plasma (mainly hydrogen atoms, protons and electrons) into the heliosphere/astrosphere.\\
The investigation of stellar magnetic activity has a long history and goes back to the 1920s when first indications of stellar flaring, but not termed as such back then, have been reported \citep{Hertzsprung1924}. Further reports followed \citep{Luyten1926, vanMaanen1940} until in the 1940s on the star L 726-8B, better known as UV~Ceti, variations of brightness in the Balmer lines were then termed as flares, in analogy to the Sun. In contrast, the first observation of a solar flare dates back to the 19th century, when \citet{Carrington1859} witnessed a white-light flare on the Sun. On stars other than the Sun, a statistical determination of flaring parameters, in dependence on spectral type, have become possible since the launch of satellite missions dedicated to the search for exoplanets using the transit method. Starting with the COnvection, ROtation and planetary Transits (CoRoT) space telescope, followed by the Kepler space telescope, hundred thousands of flares have been detected \citep{Davenport2016}. From Kepler data numerous highly energetic flares on Sun-like stars, so-called superflares, have been detected which are defined by an energy larger than 10$^{33}$~erg \citep[][and many other publications up to now]{Maehara2012}. Then, an all sky survey followed, as CoRoT and Kepler focused on fixed fields in the sky. The Transiting Exoplanet Survey Satellite (TESS) is now in its fourth year of operation, thereby having mapped the sky already twice.\\
The second highly energetic activity phenomenon, which is known from the Sun since the 1970s \citep{Tousey1973}, is the CME. CMEs are characterized by a so-called three-part-structure, namely, the core, the cavity, and the leading edge. In many cases the dense core is represented by a filament/prominence, therefore a close correlation between CMEs and filaments/prominences exists. Filaments/prominences consist of plasma captured in magnetic field lines located at a mean height of 2.6$\times$10$^{4}$~km above the solar photosphere \citep{Wang2010} and are pronounced in the Balmer lines. Filaments are seen in absorption in front of the solar disk, and are known as prominences if observed on the limb in emission\footnote{Hereafter, we will use the term ``prominence'' except where the distinction is relevant, as prominences and filaments represent the same physical phenomenon.}. From an evolutionary perspective, erupting prominences represent CMEs at an early evolutionary stage at low coronal heights \citep[cf. the CSHKP model of eruptive flares;][and references therein]{Magara1996}, although not all solar CMEs emerge from erupting filaments \citep{Gopalswamy2003}.\\
Especially since the operational start of the Large Angle Spectroscopic Coronagraph (LASCO) onboard the Solar and Heliospheric Observatory (SOHO), the Sun is monitored continuously and a statistical determination of CME parameters has become possible. On the Sun, CMEs reveal mean masses of $\sim$10$^{15}$~g with a maximum mass of 10$^{17}$~g. Solar CME speeds show an average of close to 400~km~s$^{-1}$, but may reach up to a few 1000~km~s$^{-1}$, whereas solar CME kinetic energies show a mean of 10$^{30}$~erg \citep[see ][for a statistical analysis of solar CMEs over solar cycles 23 and 24]{Lamy2019}.\\
CMEs on the Sun are known to be correlated to flares, this correlation approaches 100\% for energetic flares, the so-called X-class ($>$10$^{-4}$W~m$^{-2}$) flares \citep{Yashiro2009}. But there is also an exception to the rule, as \citet{Thalmann2015} reported on 6 confined X-class flares, i.e. flares without CMEs, occurring in October 2014 originating from one of the largest known solar active regions, indicating that magnetic confinement may play an important role in the detection of CMEs on active stars due to their larger spot sizes \citep{Shibata2013}.\\
Solar CMEs are routinely detected in coronagraph images, which is not possible with other stars, as the circumstellar environment cannot be spatially resolved by current instrumentation. Stellar CMEs are thus more difficult to detect than flares, as the latter can already be seen in integrated light and require time series photometry only. On the other hand, CMEs are not visible in integrated light in photometric time series. For the detection of CMEs other methods need to be applied. In principle one distinguishes between direct and indirect methods. The direct method is the signature of plasma moving away from a star, this is recognized as a Doppler-shifted signature in spectra in various wavelength domains, from optical to X-rays. The indirect signatures relate to phenomena which are known to be correlated with CMEs on the Sun, such as radio type II and type IV bursts, and coronal dimmings. Furthermore, there are methods which interpret variations in the hydrogen column density during flares as CME plasma obscuring active regions (continuous absorptions in X-rays), the sudden appearance of UV lines in absorptions as CMEs crossing the line-of-sight in a close binary system, dips in light curves of eclipsing binaries after a flare as CMEs, or pre-flare dips in light curves as destabilizing filaments.\\
Improving our knowledge on stellar CMEs is necessary because of several reasons. CMEs are one of the main drivers for space weather, which means that they interact with (exo-)planetary atmospheres. Depending on the magnetic field protection of a planet, the distance of the planet to the host star, and the activity level of the host star, (exo-)planetary atmospheres can be in the worst case scenario completely eroded \citep{Khodachenko2007, Lammer2007, Cohen2011, Airapetian2020} so that the origin and evolution of life would not be possible. CMEs play also a role in stellar mass- and angular momentum loss, and this is in turn relevant for stellar spin-down, and therefore also for stellar evolution. 
In the following, we review the currently applied methods to detect stellar CMEs, as well as their interpretations.

\section{\uppercase{Doppler-shifted emission/ab\-sorption signatures}}
\label{Dopplermethod}
This is the only direct method of how to detect plasma from either CMEs or prominences erupting from stars. When plasma is ejected from a star, then the moving plasma will produce a signature either on the blue (plasma moving into the observer's direction) or on the red side (plasma moving away from the observer) of the stellar spectral line. The appearance of such signatures in specific spectral lines depends on the plasma parameters, such as composition, temperature, and density. To search for CMEs using this method, spectroscopic time series are needed. These spectroscopic time series can be obtained in various wavelength domains, such as the optical, UV, or X-rays. If one focuses on solar-like stars, then, according to the solar-stellar analogy, it can be expected that signatures of CMEs on the Sun may also appear in similar wavelength domains on solar-like stars.
\subsection{At optical wavelengths}
\begin{figure*}[htp]
\centerline{\includegraphics[width=15cm, keepaspectratio]{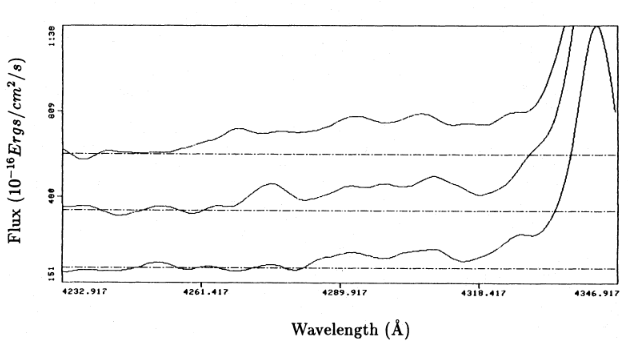}}
\caption{First detection of a blue-wing enhancement in Balmer lines interpreted as a stellar mass ejection. Shown is a sequence (from top to bottom) of smoothed H$\gamma$ residual spectra (quiescent component subtracted) of the young and active dMe star AD~Leo. One can clearly see the broad enhancement on the blue side of H$\gamma$ which slows down and loses flux as time evolves. Taken from \citet{Houdebine1990}, reproduced with permission $\copyright$ ESO.}
\label{fig1}
\end{figure*}

\begin{figure}[htp]
\vspace*{-1cm}
\centerline{\includegraphics[width=\columnwidth, keepaspectratio]{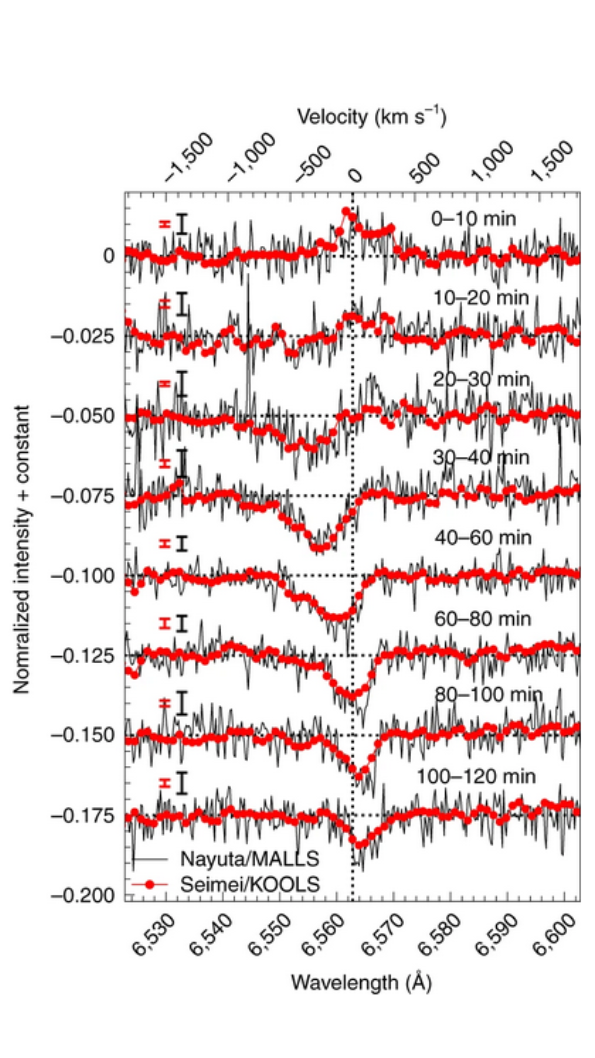}}
\vspace*{-1cm}
\caption{First detection of a blue-wing absorption in Balmer lines interpreted as a stellar mass ejection. Shown is a sequence (from top to bottom) of H$\alpha$ residual spectra (quiescent component subtracted) of the young and active solar analogue EK~Dra. One can clearly see the absorption on the blue side of H$\alpha$ which evolves from blue to red. Black solid lines and red dots represent data from two observatories having recorded the event simultaneously. The vertical dotted line marks the center of the stellar H$\alpha$ line. Taken from \citet{Namekata2022}.}
\label{fig2}
\end{figure}

The first stellar analog to a solar CME or prominence eruption was presented by \citet{Houdebine1990}. A very broad blue wing enhancement in the Balmer line H$\gamma$ was detected on the $\sim$200~Myr old dMe star AD~Leo (see Fig.~\ref{fig1}), occurring at the onset of a large complex flare \citep{Rodono1985, Rodono1989}. The measured projected maximum velocity was 5800~km~s$^{-1}$, which is exceptionally large and represents also the fastest stellar CME detected so far. The bulk velocity of the event seems to be around 3000~km~s$^{-1}$, as estimated by eye from Fig.~\ref{fig1}. A weaker counterpart of the blue wing enhancement was also detected in the H$\delta$ line. The
authors calculated the mass based on the expression of the spectral line emissivity together with population ratios derived from a non-local thermodynamic equilibrium (NLTE) atmosphere model of a dM star and found a mass of 7.7$\times$10$^{17}$~g, which is roughly an order of magnitude larger than the most massive CMEs on the Sun. This value is a lower limit as the line emissivity gives a lower limit to the number of emitting atoms. A subsequent analysis of this flare also showed indications of pre-flare motions of a dark filament, as well as oscillations in a prominence which later expanded and disrupted during the gradual phase of the flare \citep{Houdebine1993a, Houdebine1993b}.\\
\citet{Gunn1994} presented a slower event detected during a flare on the similarly aged dMe star AT~Mic, which revealed a projected bulk velocity of $\sim$250~km~s$^{-1}$. The authors interpreted this event as high-velocity chromospheric evaporation. Chromospheric evaporation is a process which is known from the Sun \citep{Canfield1990, Heinzel1994a, Li2022} to occur during the onset of flares when heated plasma is moving upwards in the flare loops. Usually the evaporating plasma is heated within seconds to minutes on the Sun and is mainly observed at soft X-ray wavelengths. 
Chromospheric evaporation in the Balmer lines is hard to detect as the plasma is heated quickly to very high temperatures, especially for stellar spectroscopic observations with integration times of the order of minutes such motions are thus unlikely to be observed in the Balmer lines (cf. Section~\ref{disc}). Although \citet{Gunn1994} interpreted these observed signatures on AT~Mic as chromospheric evaporation, the deduced parameters also allow an interpretation as eruptive prominence.\\
A few years later, \citet{Guenther1997} performed a search for flares on T-Tauri stars in the Chamaeleon association. The authors reported on one flare event revealing a distinct blue asymmetry with a projected bulk velocity of 600~km~s$^{-1}$ and an estimated mass range of 0.2-7.8$\times$10$^{19}$~g. The event was detected on DZ~Cha, a weak-line T-Tauri star (WTTS) of spectral type M.\\
\citet{Vida2016} presented an investigation of the activity of the fast rotating, fully convective dMe star V374~Peg, in which both photometric and spectroscopic data were analyzed. The spectral data revealed several flares, one of them being a complex flare event with several blue wing enhancements. The fastest event had a projected bulk velocity of $\sim$300~km~s$^{-1}$ and a mass of $\sim$10$^{16}$~g, and was interpreted as a mass ejection, whereas the several low-velocity blue wing enhancements were interpreted as failed eruptions. The ejection scenario lasted for half an hour, followed by a longer low-velocity red asymmetry lasting for about one hour, interpreted as back-flowing material, reminiscent of solar observations \citep[e.g.][]{Christian2015}. The blue wing enhancements were visible from H$\alpha$ throughout H$\delta$.\\
Only recently, \citet{Namekata2022} presented spectroscopic time series of the young solar analogue EK~Dra (see Fig.~\ref{fig2}). For the first time an absorption feature on the blue side of H$\alpha$ was detected during a superflare on a main-sequence star. The projected bulk velocity of this feature was $\sim$510~km~s$^{-1}$, which is close to the escape velocity of EK~Dra. The absorption feature appeared first at maximum velocity and then reduced its velocity until it reached zero speed, then changed its direction indicated by its appearance on the red side of H$\alpha$ with low velocity. This dynamic behaviour indicates that some mass was falling back towards EK~Dra.\\
A number of further studies dedicated to specific stars reported the detection of blue wing asymmetries during  flares with projected bulk velocities well below the escape velocities of the stars. \citet{Lopez-Santiago2003} reported on a blue wing emission feature on the active K 
dwarf PW~And, which the authors assigned to possible dynamic processes taking place during a flare; \citet{FuhrmeisterSchmitt2004} found in H$\alpha$ and H$\beta$ spectra of the dM9 star DENIS~104814.7-395606.1 a blue wing asymmetry during a flare with a projected bulk velocity of $\sim$100~km~s$^{-1}$ for which the authors favoured a mass ejection scenario; \citet{Hill2017} investigated spectropolarimetric data of two WTTS and found red wing absorptions and blue wing emissions which the authors assigned to infall of plasma along the flare loops (red) and possibly originating from erupted plasma (blue); \citet{Honda2018} reported on blue wing asymmetries during a flare on the young and active dMe star EV~Lac, the authors discussed several scenarii from flare related processes to filament activation; \citet{Muheki2020a, Muheki2020b} analyzed spectroscopic time series of AD~Leo and EV~Lac and found a few dozens of flares and blue wing enhancements with low projected velocities, except for one case reaching 220~km~s$^{-1}$ which the authors suspected to originate from an erupting prominence; \citet{Maehara2021} reported on blue wing enhancements on the young and active dMe star YZ~CMi possibly originating from prominence eruptions with deduced projected velocities in the range of 80-100~km~s$^{-1}$ and masses in the range of 10$^{16}$-10$^{18}$~g; \citet{Johnson2021} reported on a possible failed eruption on the young dMe star GJ~3270; \citet{Wang2021} found during flares on two M dwarfs blue wing enhancements with low bulk velocities, but with maximum velocities exceeding the stars' escape velocities, which the authors attributed to CMEs with deduced masses in the range of 10$^{18}$-10$^{19}$~g; and finally, \citet{Wang2022} investigated spectroscopic time series of two M dwarfs and found very broad H$\alpha$ profiles during flares with maximum velocities in the range of 700-800~km~s$^{-1}$, but bulk velocities close to 0~km~s$^{-1}$, for which the authors discussed both Stark broadening from the flare, as well as a scenario of limb CMEs (with estimated masses of 10$^{18}$-10$^{19}$~g) as possible interpretations.\\
Another approach using the method of Doppler shifted 
\begin{figure*}[htp]
\vspace*{-0.5cm}
\centerline{\includegraphics[width=16cm, keepaspectratio]{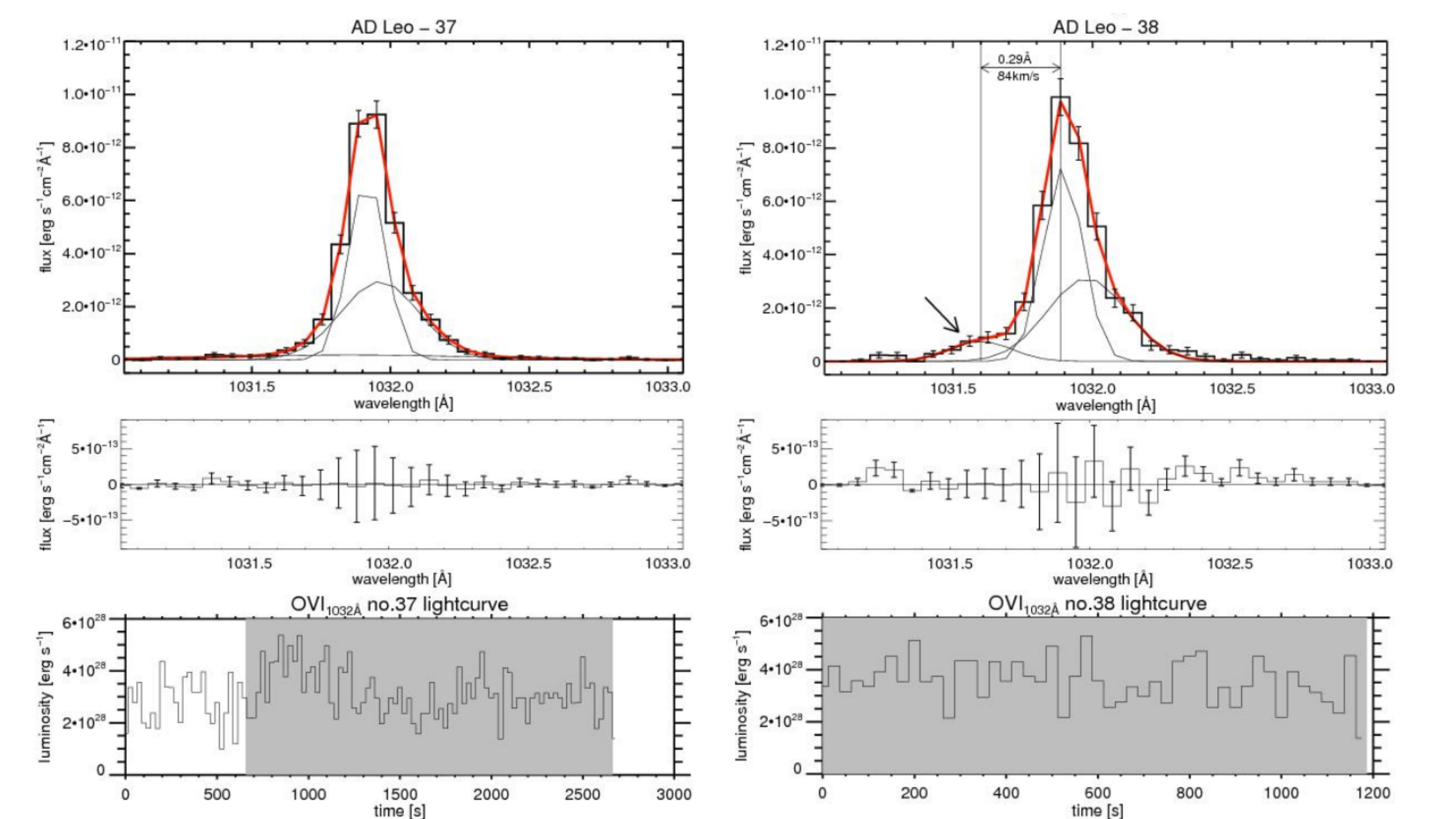}
}
\vspace*{-0.0cm}
\caption{O\,VI spectral profiles, residuals, and corresponding light curves of a flare (left column) and the successive spectrum (right column) showing a blue-wing enhancement at a projected velocity of 84~km~s$^{-1}$ on AD~Leo. Taken from \citet{Leitzinger2011a}, reproduced with permission $\copyright$ ESO.}
\label{fig2a}
\end{figure*}
emission/absorption is, apart from obtaining spectroscopic time series of one star, the usage of data archives, thereby investigating a larger number of stars. Usually archival data are obtained for different science cases, thereby not necessarily having the desired observational parameters (signal-to-noise, exposure time, wavelength coverage, duration, etc.) but the data can still be used to search for stellar CMEs. \citet{Fuhrmeister2018} searched for Balmer line variability in spectroscopic time series of the ``Calar Alto high-Resolution search for M dwarfs with Exoearths with Near-infrared and optical Echelle Spectrographs'' (CARMENES). These authors found few dozens of blue/red wing enhancements on M dwarfs, all well below the escape velocities of the stars.
\citet{Vida2019} used the Polarbase archive, hosting data from the Narval spectrograph installed at T\'elescope Bernard Lyot at Pic du Midi and from the ``Echelle SpectroPolarimetric Device for the Observation of Stars'' (ESPaDOnS) installed at the Canada-France-Hawaii Telescope (CFHT) on Hawaii, to search for CMEs on single M dwarfs. Dozens of blue wing asymmetries during flares in the Balmer lines were detected, the fastest ones being in the order of a few hundreds of km~s$^{-1}$. The majority of events showed low projected velocities. \citet{Koller2021} used data from the Sloan Digital Sky Survey (SDSS) data release 14 to search for CMEs. The search included spectral types F-M and revealed few events with blue and red wing enhancements, all occurring on dMe stars. \citet{Lu2022} utilized data from the Large Sky Area Multi-Object Fibre Spectroscopic Telescope (LAMOST) to search for CMEs, also 
here only a handful of events were found which showed red and blue wing enhancements in the Balmer lines, but in one case also additional blue wing enhancements in the chromospheric Mg\,I triplet lines.\\
Apart from the many studies reporting on blue wing asymmetries with low projected velocities, there were also studies solely dedicated to the search for CMEs reporting on non-detections. \citet{Leitzinger2014} presented multi-object spectroscopic observations of members of the southern open cluster Blanco-1. Except for a few flares (one with symmetrically broadened wings), no signatures of CMEs could be found, which may be explained by the rather short on-source time of 5~hours, although a few dozens of stars were observed simultaneously. \citet{Korhonen2017} reported on further multi-object spectroscopic observations of open cluster member stars in IC2391, NGC2516, NGC3532, h~Per, and IC348. Only on one star in IC348 H$\alpha$ variability was detected. \citet{Leitzinger2020} used the phase 3 archive of the European Southern Observatory (ESO) of the High Accuracy Radial velocity Planet Searcher (HARPS), as well as the Polarbase archive, to search for CMEs on solar-like stars, including spectral types F,G, and K. In more than 3700 hours of on-source time no signatures of CMEs could be found. This shows that the detection of CMEs on main-sequence stars other than M-type is far more unlikely.  
\subsection{At UV/X-ray wavelengths}
The number of publications on stellar CME studies at shorter wavelengths is much lower than in the optical, because satellite observations, especially longer time series, are not so abundantly available as for groundbased observations.\\
\citet{Leitzinger2011a} searched for signatures of stellar CMEs in spectroscopic data of the Far Ultraviolet Spectroscopic Explorer (FUSE). Only one blue wing enhancement could be found on AD~Leo occurring in one spectrum after a flare in the first component of the O\,VI duplet (see Fig.~\ref{fig2a}). In solar CMEs, the O\,VI duplet is observed by the UltraViolet Coronagraph Spectrometer (UVCS) onboard SOHO, an instrument which has detected more than thousand solar CMEs. The projected velocity of this event was low 
(84~km~s$^{-1}$).\\ \citet{Argiroffi2019} presented Chandra X-ray observations of the magnetically active giant star HR~9024 (G1\,III). The authors found up- and downward motions in the flare loop with velocities of 100-400~km~s$^{-1}$. After the flare the authors found a blue-shift in the O\,VIII line with a projected velocity of $\sim$90~km~s$^{-1}$ which the authors ascribe to a CME. The CME event presented in \citet{Argiroffi2019} shows strong similarities to the event presented in \citet{Leitzinger2011a} with respect to timing (occurrence of a blue-shifted emission after the flare), velocity, and the element in which the event was detected \citep[cf. Fig.~\ref{fig2a} and Fig.~2f in][]{Argiroffi2019}. The durations of the events differ strongly, 
\begin{figure}[htp]
\vspace*{-0.1cm}
\centerline{\includegraphics[width=\columnwidth, keepaspectratio]{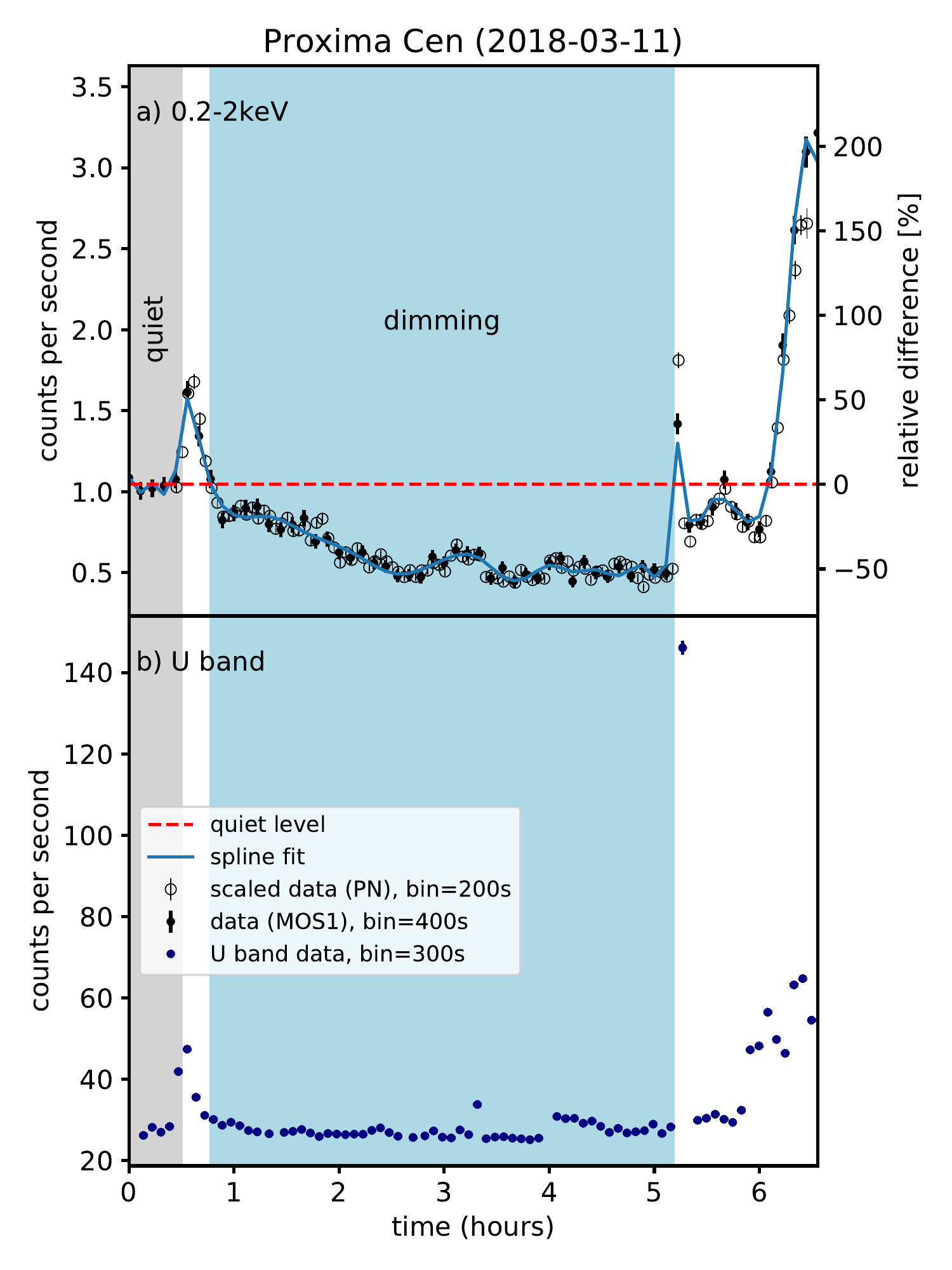}}
\vspace*{-0.1cm}
\caption{XMM-Newton X-ray (upper panel) and U-Band (lower panel) light curves of Proxima Cen showing three flares. After the first flare a long-lasting dimming is evident. The red dashed line marks the pre-flare level of Proxima Cen in the 0.2-2~keV band. This is a representative event out of 21 events. Adapted from \citet{Veronig2021}.}
\label{fig3}
\end{figure}
as the event on AD Leo (both flare and asymmetry) covers $\sim$1~hour and the event on HR~9204 covers $\sim$25~hours.\\
\citet{Bourrier2017} performed Ly$\alpha$ observations of the TRAPPIST-1 exoplanetary system to search for extended planetary hydrogen atmospheres. The authors compared Ly$\alpha$ flux levels obtained at different epochs of transits of TRAPPIST-1c and recognized an absorption in one epoch which was suspected to originate from an extended hydrogen envelope of TRAPPIST-1c. But this lower flux in one epoch was not evident in the following epoch, therefore the authors suspected, beside other scenarii, a stellar filament eruption which might have caused the transient lower flux in Ly$\alpha$.\\
\citet{Chen2022} analyzed Chandra observations and found plasma up- and downflows with velocities of a few tens up to 130 km~s$^{-1}$ during several flares on EV~Lac, comparable to the findings presented in \citet{Argiroffi2019} on the giant star HR~9024. Also cool to warm upflows were detected in one flare together with a decreasing plasma density which the authors attributed to a possible filament eruption. 

\section{\uppercase{Coronal dimmings}}
\label{dimmingmethod}
On the Sun, coronal dimmings were identified in images in the soft X-ray and EUV ranges \citep[e.g.][]{Hudson1996, Mason2014, Dissauer2018}. They are frequently temporally and spatially associated with the eruption of CMEs. 
The signature of these dimmings typically appears in cool coronal lines, whereas a flare associated with the CME is typically more pronounced in hot lines (although some 
lines show signatures of both, flare and dimming). Dimming signatures can also be detected in the light curves of disk-integrated data, where they appear as a drop below the pre-flare level after a flare event, reaching a minimum, and then slowly recovering to the pre-flare flux. This motivates the search for these signatures on stars other than the Sun. In fact, \citet{Harra2016} identified coronal dimmings as the only reliable signature for the occurrence of a flare-associated CME from a pool of flare- and active region properties in solar X-class flares. Coronal dimmings are interpreted as regions of expelled coronal plasma due to CMEs, which gradually recover to normal coronal levels within hours to days. Several studies have found correlations between dimming parameters and the properties of associated CMEs; the dimming depth is related to the CME mass, whereas the slope of the dimming decrease is related to its speed \citep{Mason2016, Jin2022}. However, the EUV range which encompasses most of the prominent dimming lines on the Sun is difficult to observe for other stars because of absorption by the interstellar medium. Models predict that in the hotter coronae of active stars dimming signatures may also be common in the soft X-ray range \citep{Jin2020}. In a recent study by \citet{Veronig2021} stellar X-ray and EUV archival data were searched for light curves with morphologies similar to the solar disk-integrated data (see Fig.~\ref{fig3}). They found 21 dimming signatures in X-ray/EUV light curves on 13 G-, K- and M-type pre-main-sequence and main-sequence stars following flare events. The maximum dimming depths of 5-56\% are about an order of magnitude larger than on the Sun, however, typical solar-like dimming depths of a few percent would not easily be detectable in stellar observations due to higher noise levels in the data and larger intrinsic variability of active stars. The rise times (i.e. dimming start to maximum depth) were with about 2\,h similar to the Sun. The total durations and recovery times were on average shorter than on the Sun, but we note that most of the stellar observations were not long enough to observe the return to pre-flare levels and/or the dimming was overlapping with a subsequent flare event so that the true recovery time could not be obtained. Recently, \citet{Loyd2022} attempted to detect dimmings in FUV lines of the active K dwarf $\epsilon$~Eri, but found no significant dimmings following the three flares they observed. They therefore provided upper limits on the masses of potentially associated CMEs in the plasma temperature range probed by the studied emission lines.\\

\section{\uppercase{Radio bursts}}
\label{radiomethod}
On the Sun, radio bursts have been recognized to be correlated to other solar phenomena. Radio type II bursts (bursts showing a slow frequency drift in dynamic spectra) are a signature of a shock wave, as electrons are accelerated at the shockfront \citep{Holman1983}. As CMEs can drive shock waves, a correlation between both phenomena exists in the hecto-/decameter regime on the Sun \citep[e.g.][]{Reiner2001a}. Here every type II bursts was correlated with a CME, but not every CME was correlated with a type II burst. In the meter regime, \citet{Classen2002} found that one third of the investigated CMEs was associated with type II bursts excited from the leading edge of the CME, one third with flare related shocks, and one third with shocks related to the flanks or inner parts of the CME. Type III bursts (bursts showing a fast frequency drift in dynamic spectra) are usually related to flaring regions from which relativistic electron beams originate \citep[see e.g.][]{Bastian1998}. Type IV bursts (broad-band continuum bursts) are often related to CME flux ropes or foot points of magnetic loops \citep[e.g.][]{Salas2020}. Their correlation with CMEs is not as well established as the one for type II bursts, but has been subject of several recent studies \citep[e.g.][]{Kumari2021, Morosan2021}.\\
On flare stars, the investigation of radio bursts dates back to the 1960s \citep{Lovell1963, Slee1963}, and since then many prominent flare stars have been observed in various radio frequency regimes \citep[see e.g.][]{Bastian1990, Bookbinder1991, Guedel2002}.
\begin{figure*}[htp]
\vspace*{-0.1cm}
\centerline{\includegraphics[width=\textwidth, keepaspectratio]{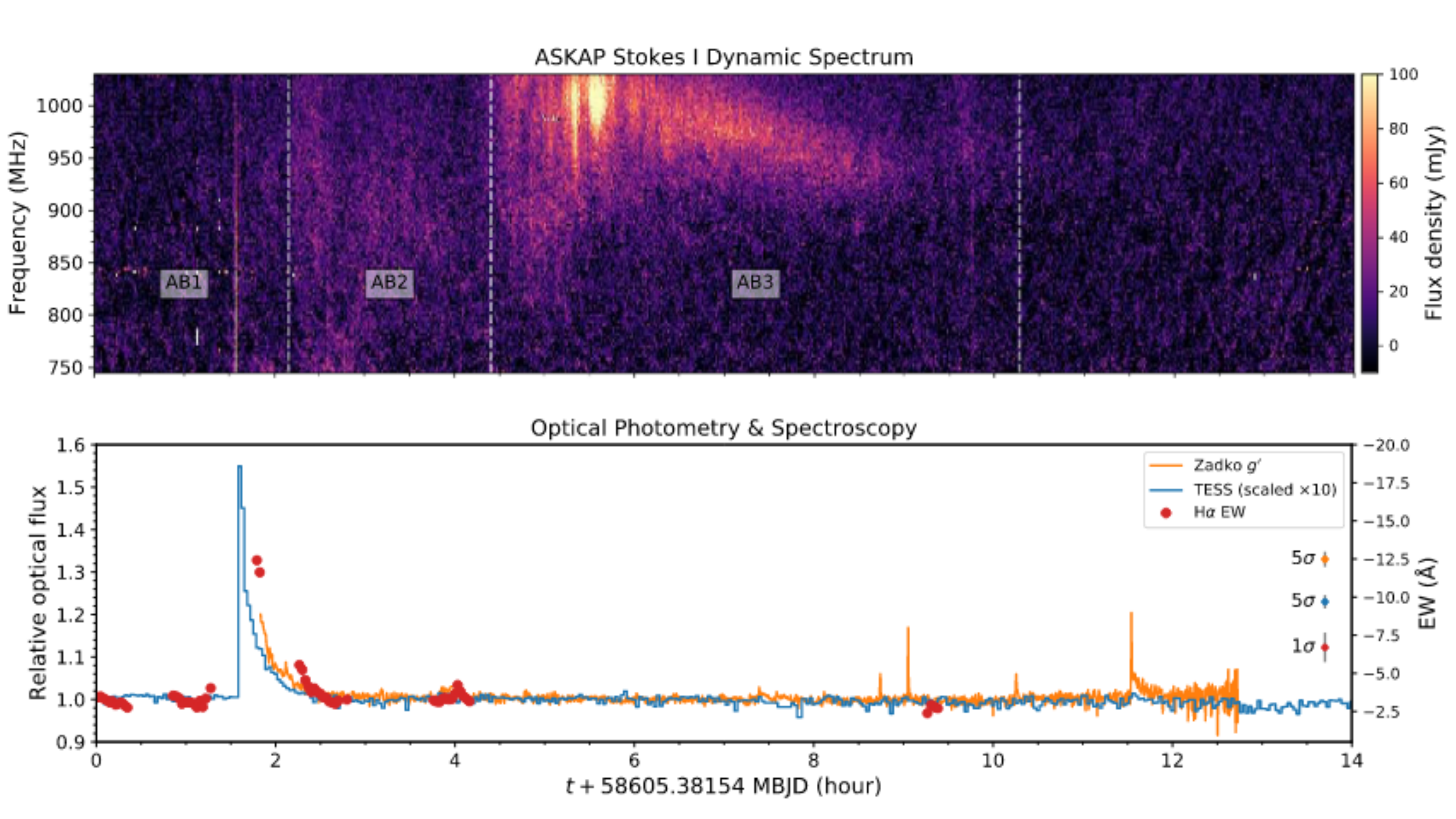}}
\vspace*{-0.1cm}
\caption{Dynamic spectrum (upper panel) of stellar radio bursts on Proxima Cen. In the lower panel, the light curve during the same time range as the dynamic spectrum is shown and reveals that type IV-like radio bursts (AB2, AB3) occurred during and after the decay phase of the optical flare. Taken from \citet{Zic2020}. $\copyright$ AAS. Reproduced with permission.}
\label{fig4}
\end{figure*}
\citet{Slee1963} found a strong radio burst on the dM3e star V371~Ori which was also observed coordinated in the optical. These radio observations were performed at the Mills Cross aerial near Sydney and Parkes, New South Wales, Australia. Although observing at three discrete frequencies, the full light curve could be observed only at 410~MHz. \citet{Jackson1990} reported on meter/decameter observations of twelve prominent flare stars (incl. AD~Leo, AU~Mic, YZ~CMi etc.) using the Clark Lake radio telescope. In more than 140 hours of observations no radio bursts could be detected except two marginal events at the highest frequency of 110.6~MHz. In the 1990s there were several observing campaigns to search for stellar activity in form of radio bursts using the World's largest decameter radio array, the Ukrainian T-shaped Radio telescope second modification (UTR-2), operating in the frequency domain of 8-32~MHz. \citet{AbdulAziz1995} reported on a coordinated observing campaign of EV~Lac in 1992 using the UTR-2, the Jodrell Bank radio interferometer operating at centimeter wavelengths, and optical photometry from several observing sites. The authors found numerous optical flares from EV~Lac, and also some radio bursts occurring in the same time frame. Although discrimination techniques \citep{Abranin1994} were applied to the bursts, there was still some doubt that the rather short-lived (in the range of few seconds) bursts were indeed originating from EV~Lac, especially as no centimeter emission was detected during noticeable decameter bursts. One has to keep in mind that decametric radio arrays have limited spatial resolution and sensitivity, which means that only strong bursts can be detected. A similar observing campaign dedicated to EV~Lac was described in \citet{Abranin1997}, but this time no radio bursts were identified which could be of stellar origin. One more campaign was undertaken by \citet{Abranin1998}, again focusing on EV~Lac. This time, 18 radio bursts were detected, from which one burst satisfied most discrimination criteria and was observed simultaneously with an optical flare.\\
\citet{Leitzinger2009} reported on a similar approach to \citet{AbdulAziz1995} and \citet{Abranin1997, Abranin1998}, this time focusing on the two similarly active M dwarfs AD~Leo and EV~Lac. A main difference to the prior observing campaigns was the scientific goal, which shifted from the general detection of possible stellar radio bursts to the detection of stellar analogues of the solar radio type II bursts as CME signatures. Therefore, it was necessary to involve multi-channel receivers to obtain frequency information with a certain resolution across the observable frequency range, enabling the detection of drifting bursts. Such receivers were available at the UTR-2 at that time with 1024 and 2048 channels, respectively. The campaigns revealed a number of bursts, but all of them were fast-drifting and therefore not type II-like.
\citet{Boiko2012} continued the flare star observations  using the UTR-2 and found more than 200 radio bursts on EV~Lac and AD~Leo in observations from 2010/2011 with parameters more similar to solar type III bursts. \citet{Konovalenko2012} reported on AD~Leo observations with the UTR-2 in 2011 and presented bursts with a high probability of being of stellar origin by applying advanced techniques to discriminate between stellar and non-stellar origin. \citet{Crosley2016} utilized the Low Frequency Array (LOFAR), operating at similar frequencies as the UTR-2, to search for type II emission from YZ~CMi. In 15 hours of observations no type II-like bursts could be found.\\
The search for stellar analogues of solar type II emission continued at higher frequencies. \citet{Crosley2018b, Crosley2018a} used the Jansky Very Large Array (JVLA) to monitor the young and active M-dwarf binary EQ~Peg at frequencies of 230-470~MHz. In a total observing time of 64 hours, two bursts were detected which the authors identified to be not type II-like, as the expected parameters did not match the observed ones. \citet{Crosley2018b} concluded that they doubt that a high flaring rate of a star also means a high CME rate. \citet{Villadsen2019} also used the JVLA in the frequency bands 224–482~MHz and 1–6~GHz to search for radio bursts in five active dMe stars and detected nineteen coherent radio bursts with different morphology, different duration, but all with a high degree of circular polarization. The authors stated furthermore that none of the bursts resemble solar type II bursts, but they did not rule out the occurrence of CMEs on the investigated stars, as some CMEs may be radio-quiet. \citet{Mullan2019} hypothesized that, because of the strong magnetic fields in M dwarfs, the CME velocity would have to be unrealistically high to produce type II emission, thereby supporting the radio-quiet CME scenario proposed by \citet{Villadsen2019}. \citet{Zic2020} performed a multi-wavelength campaign of our next stellar neighbour, Proxima Centauri, a dM5.5e star of solar age. Using optical photometry and spectroscopy from TESS and several Australian observatories, as well as radio observations from the Australian Square Kilometre Array Pathfinder (ASKAP) using a central frequency of 888~MHz with a bandwidth of 288~MHz, the authors found during a long-lasting optical flare a sequence of coherent radio bursts from which the trailing long-duration burst was identified to be type IV-like (see Fig.~\ref{fig4}) and indicative of a CME. In an earlier study, \citet{Kahler1982} claimed the detection of a type IV-like burst, starting 15~minutes after a flare, on YZ CMi at 408~MHz (Jodrell Bank), using several radio and optical (photometry and spectroscopy) observing facilities.

\section{\uppercase{Continuous absorptions in\\ X-rays}}
\label{contabsmethod}
\begin{figure}[htp]
\vspace*{-0.1cm}
\centerline{\includegraphics[width=\columnwidth, keepaspectratio]{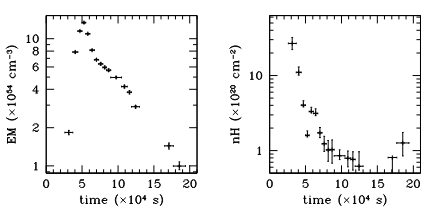}}
\vspace*{-0.1cm}
\caption{Emission measure (left panel) and hydrogen column density (right panel) of a flare on the eclipsing binary Algol. The large column density and its long decay is explained in terms of moving, cool absorbing material in the line-of-sight, i.e. a CME. Taken from \citet{Favata1999}, reproduced with permission $\copyright$ ESO.}
\label{fig5}
\end{figure}

During some stellar flare events, indications of transient absorptions were found in X-ray or UV observations, which may be interpreted as erupting filaments or CMEs moving across, and thereby temporarily obscuring, the flaring region. \citet{Haisch1983} detected an increased hydrogen column density lasting a few minutes during an X-ray flare on Proxima Centauri and suggested the passage of an erupting prominence as a potential interpretation. An X-ray flare on V773~Tau also showed an increased hydrogen column density during the flare peak, decreasing during the decay phase \citep{Tsuboi1998}. A similar behaviour was observed in a large X-ray flare on Algol \citep[see Fig.~\ref{fig5},][]{Favata1999} and was interpreted as possible absorption by a flare-associated CME which could provide the absorbing material. This event was later reanalyzed by \citet{Moschou2017} who applied a geometric model of a CME to constrain possible physical parameters of this event. They found good agreement of the column density evolution with a self-similarly expanding CME propagating with approximately constant speed, and estimated a mass in the 
order of $10^{21}{-}10^{22}$\,g. Similar increases in hydrogen column density during flares have been reported in the literature \citep[e.g.][]{Ottmann1996, Franciosini2001, Briggs2003, 
Pandey2012} and are summarized in \citet{Moschou2019}, who used their geometric model \citep{Moschou2017} to estimate physical parameters of all these events. The derived CME masses and velocities range from about $10^{16}$ to $10^{23}$\,g, and few tens to few thousands of km\,s$^{-1}$, respectively. Although most of such flare-related absorption signatures were detected in X-rays, there is one report presenting analogous findings in UV data. After a strong flare on the active M dwarf EV~Lac, several UV line fluxes dropped by a factor of two for about 1.5\,h, which was interpreted as obscuration by cool material like from an erupting prominence \citep{Ambruster1986}.
\section{OTHER METHODS}
\label{othermethods}
\begin{figure}[htp]
\vspace*{-0.1cm}
\centerline{\includegraphics[width=\columnwidth, keepaspectratio]{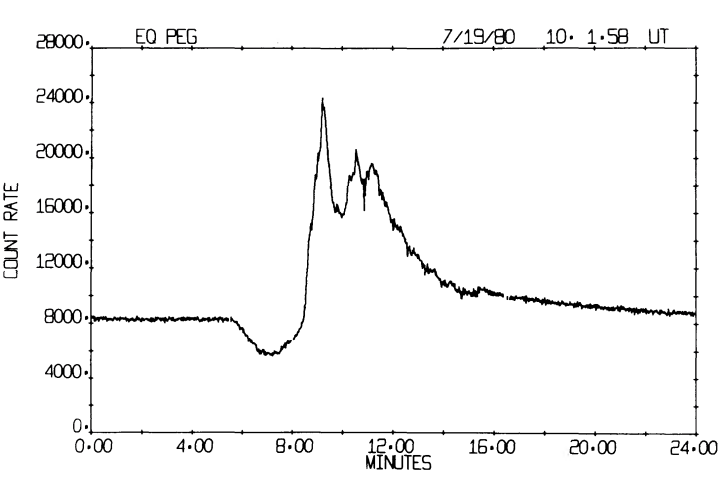}}
\vspace*{-0.1cm}
\caption{U-band light curve of a flare on EQ~Peg showing a pronounced dip shortly before the impulsive phase of the flare. Taken from \citet{Giampapa1982}. $\copyright$ AAS. Reproduced with permission.}
\label{fig6}
\end{figure}

Beside all the aforementioned publications using the different methodologies to detect stellar CMEs, there are also other signatures which have been interpreted as possible CMEs in the literature.\\
Several authors \citep{Cristaldi1971, Deming1972, Cristaldi1973, Flesch1974, Rodono1979, Cristaldi1980,  Mahmoud1980, Giampapa1982, Andersen1983, Doyle1988, Peres1993, Ventura1995, Leitzinger2014} have detected depressions in light curves prior to flares (see Fig.~\ref{fig6}). These ''dips`` have durations of a few minutes and depression depths from a few and up to 25\%. It is still not well understood what the cause of these pre-flare dips is. \citet{Grinin1976} suggested that H$^{-}$ ions temporarily block emergent radiation, which might explain the dips. This model predicts time scales of seconds, but the observed dips usually last longer (minutes up to half an hour). \citet{Giampapa1982} suggested that the 
destabilization of filaments may be the cause for the observed dips. \citet{Leitzinger2014} constructed H$\alpha$ light curves from spectra of dMe stars and found that the dips prior to flares were caused by blue wing absorptions, but those were present only in one spectrum, i.e. the 
\begin{figure*}[htp]
\vspace*{-0.1cm}
\centerline{\includegraphics[width=\textwidth, keepaspectratio]{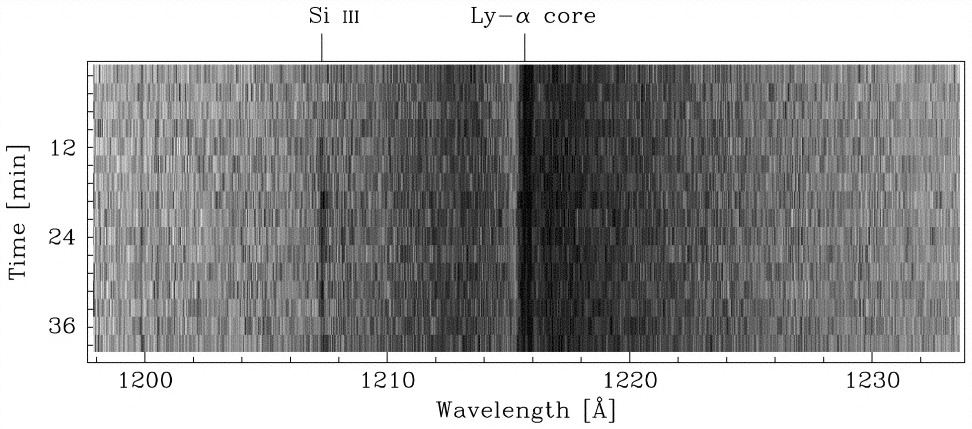}}
\vspace*{-0.1cm}
\caption{Dynamic spectrum of the pre-cataclysmic system V471~Tau. The Ly$\alpha$ spectral line is clearly visible as a broad absorption line. On the blue side of Ly$\alpha$ a vertical dark feature is visible for about half an hour, which is the Si\,III line suddenly appearing in absorption, interpreted by the authors to be caused by a CME from the K dwarf passing in front of the white dwarf. Taken from \citet{Bond2001}. $\copyright$ AAS. Reproduced with permission.}
\label{fig7}
\end{figure*}
spectrum of the dip. The authors concluded that a 
filament destabilization should be seen in more than one spectrum and suggested that the observed dips in H$\alpha$ are thus more likely linked to pre-flare processes.\\
\citet{Bond2001} investigated the pre-cataclysmic binary system V471~Tau, consisting of a white and a red dwarf (K2), using spectroscopic observations from the Goddard High Resolution Spectrograph (GHRS) on the Hubble Space Telescope (HST). The authors detected two transient absorption events in the Si\,III line at 1206\AA{} (see Fig.~\ref{fig7}). These transient events were interpreted by the authors to be CMEs from the red dwarf crossing the line-of-sight to the white dwarf. The authors confirmed their interpretation by the fact that the deduced parameters of the events lie in the range of solar CME parameters. Furthermore, a CME rate of 100-500 day$^{-1}$ was derived based on geometrical arguments. Already \citet{Mullan1989} reported on absorptions in various spectral lines of different ionization stages on V471~Tau 
and attributed those to a cool wind. \citet{Wheatley1998} reported also on absorption features detected in the light curve of the white dwarf, deduced from X-ray observations of the Roentgen Satellit (ROSAT). The deduced volume and mass were similar to those of solar CMEs. \citet{Parsons2013} searched for new eclipsing binaries consisting of a white dwarf and a main-sequence star, which they identified using SDSS data in Catalina Sky Survey (CSS) light curves. In one of the systems (SDSS J1021+1744), a distinct dip in the light curve was seen fifteen minutes after the eclipse. There was also evidence for a flare from the main-sequence star. The authors suggested the possibility of a CME passing in front of the white dwarf. In this case, the ejection must have been massive, as half of the light from the white dwarf was blocked to cause the observed dip. \citet{Irawati2016} presented a dozen of light curves of SDSS J1021+1744 and found recurring dips which the authors interpreted as possible prominences, which supports the interpretation of \citet{Parsons2013}.\\
\citet{Palumbo2022} presented TESS light curves of the young M dwarf TIC~234284556. The light curves revealed transit-like dips with varying duration and depth. Moreover, on a 1-day time scale the dips disappeared and reappeared. Interestingly, prior to the disappearance of the dip a flare-like event was recognised. The authors discussed a number of possible scenarii, from planets to slingshot prominences, as well as magnetospheric clouds and centrifugal breakout, which the authors favour for interpreting these dips. Magnetospheric clouds are plasma captured in stellar magnetic field lines, reminiscent of prominences, but include dust grains as well, which are able to absorb broadband stellar emission. 

\section{\uppercase{Theoretical approaches and modelling}}
\label{modelling}
Numerous theoretical studies have emerged either trying to infer CME characteristics of stars based on related phenomena, or to physically model CMEs and their signatures in the environment of other stars.

\subsection{Theoretical CME rates based on stellar properties}
Several studies aimed at the inference of stellar CME activity based on flare statistics. Flares are observed much more commonly on stars and are easier to detect than CMEs, and since flares and CMEs are closely related on the Sun, one may assume that strong stellar flares could be frequently accompanied by CMEs. \citet{Aarnio2012} combined a relationship between X-ray flare energy and CME mass from the Sun \citep{Aarnio2011} with X-ray flare statistics of pre-main-sequence stars in the Orion Nebula Cluster. Assuming that all of their powerful flares are accompanied by CMEs, they estimated mass-loss rates by CMEs 1-5 
orders of magnitude higher than the present Sun's total mass-loss rate \citep[$\dot{M}_\odot{=}2{-}3{\times}10^{-14}\,M_\odot$\,yr$^{-1}$;][]{Wang1998}. \citet{Drake2013} used a similar approach 
to estimate CME rates of main-sequence stars, but instead of using observed flare rates they scaled theoretical distributions 
according to the star's emitted X-ray luminosity. In addition, they took into account the solar flare-CME association rate \citep{Yashiro2009}, as the association increases with flare magnitude and reaches 100\% only for the strongest solar flares. They found that mass-loss rates by CMEs 
increase with stellar X-ray luminosity, reaching up to four orders of magnitude above $\dot{M}_\odot$ for the most active stars. However, the corresponding kinetic energy losses would result in unrealistically high energy requirements, thus \citet{Drake2013} suggested that there must be limitations in the simple extrapolation of solar relationships to active stars. 
\begin{figure}[htp]
\vspace*{-0.1cm}
\centerline{\includegraphics[width=\columnwidth, keepaspectratio]{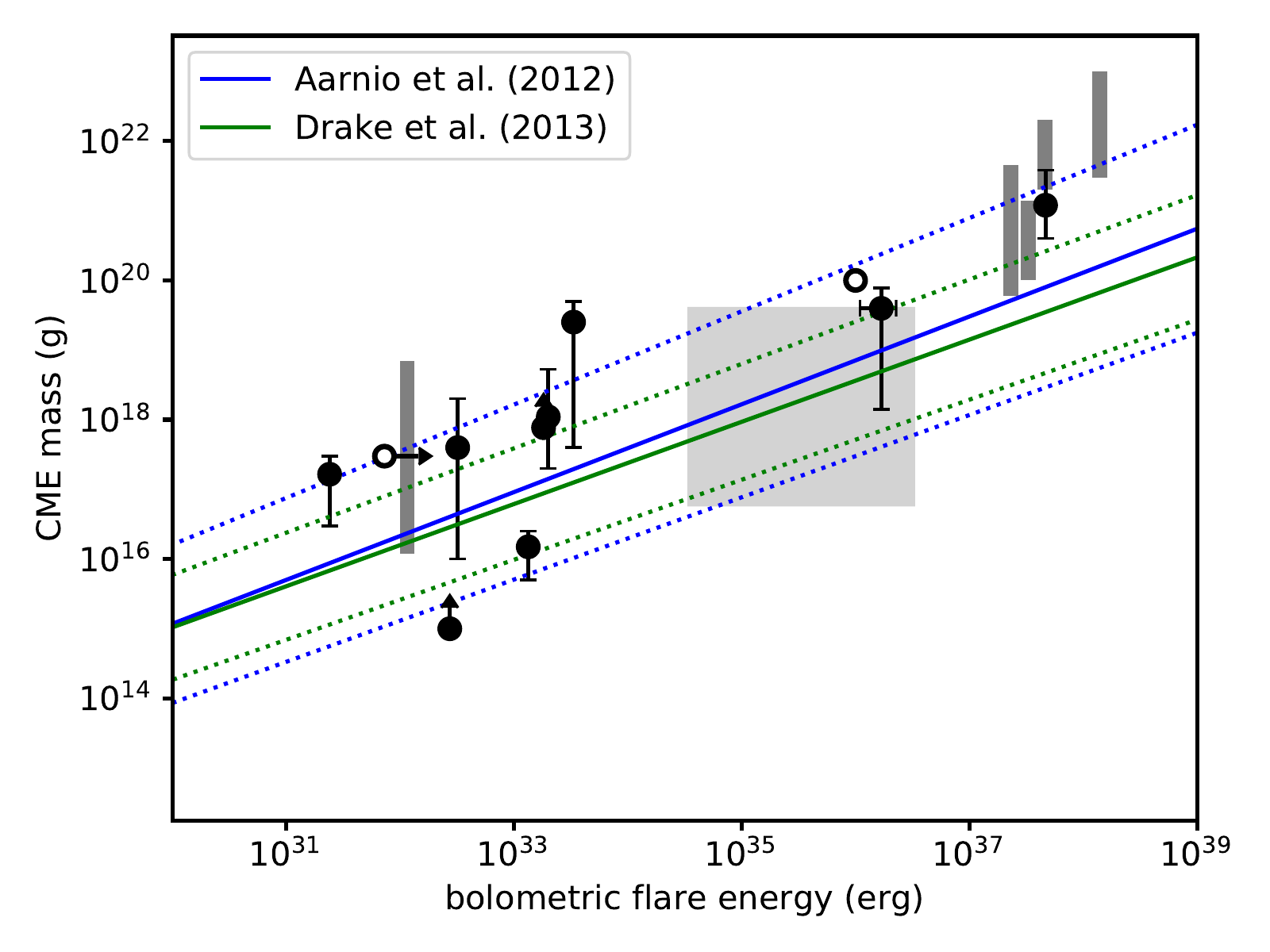}}
\vspace*{-0.1cm}
\caption{Comparison of flare energy and mass of associated CMEs. Blue and green lines show the solar relationships (solid) with their $1\sigma$ ranges (dotted) taken from \citet{Aarnio2012} and \citet{Drake2013}, respectively. Black symbols show the parameters of stellar CMEs observed with the Doppler method \citep[taken from][]{Odert2017, Moschou2019}. Errorbars were omitted if they were not given in the adopted sources, and arrows indicate lower limits. Empty symbols indicate more uncertain events from pre-flare dips \citep{Doyle1988} and an EUV flare model \citep{Cully1994}. Gray bars indicate events from X-ray absorption discussed in \citet{Moschou2019}. The lightgray-shaded area indicates typical mass ranges of stellar prominences, as well as the energy range of strong flares observed on these stars, for comparison. All flare energies have been converted from the original observations to bolometric energy using the conversion factors from \citet{Osten2015}. One can see that CME masses of active stars generally follow the extrapolated solar relationships.}
\label{figflarecme}
\end{figure}
However, the relationship between flare energy and CME mass from the Sun seems to hold also for more active stars (see Fig.~\ref{figflarecme}). \citet{Odert2017} advanced this method by estimating the stellar flare occurrence with an empirical scaling law relating stellar X-ray luminosity and flare rates \citep{Audard2000}. Their obtained CME mass-loss rates are 1-2 orders of magnitude lower for the most active stars compared to \citet{Drake2013}, i.e. only 2-3 orders above $\dot{M}_\odot$. Moreover, they checked their findings by comparing the resulting CME mass-loss rates with observed stellar mass-loss rates \citep[e.g.][and references therein]{Wood2018}, as CME-induced mass-loss would need to be lower than the total mass-loss (i.e., CMEs plus stellar wind). The extrapolated CME mass-loss rates are comparable to or lower than the observed stellar mass-loss rates for low-activity stars, but still much higher for the high-activity stars, as the observed mass-loss rates seemed 
\begin{figure}[htp]
\vspace*{-0.1cm}
\centerline{\includegraphics[width=\columnwidth, keepaspectratio]{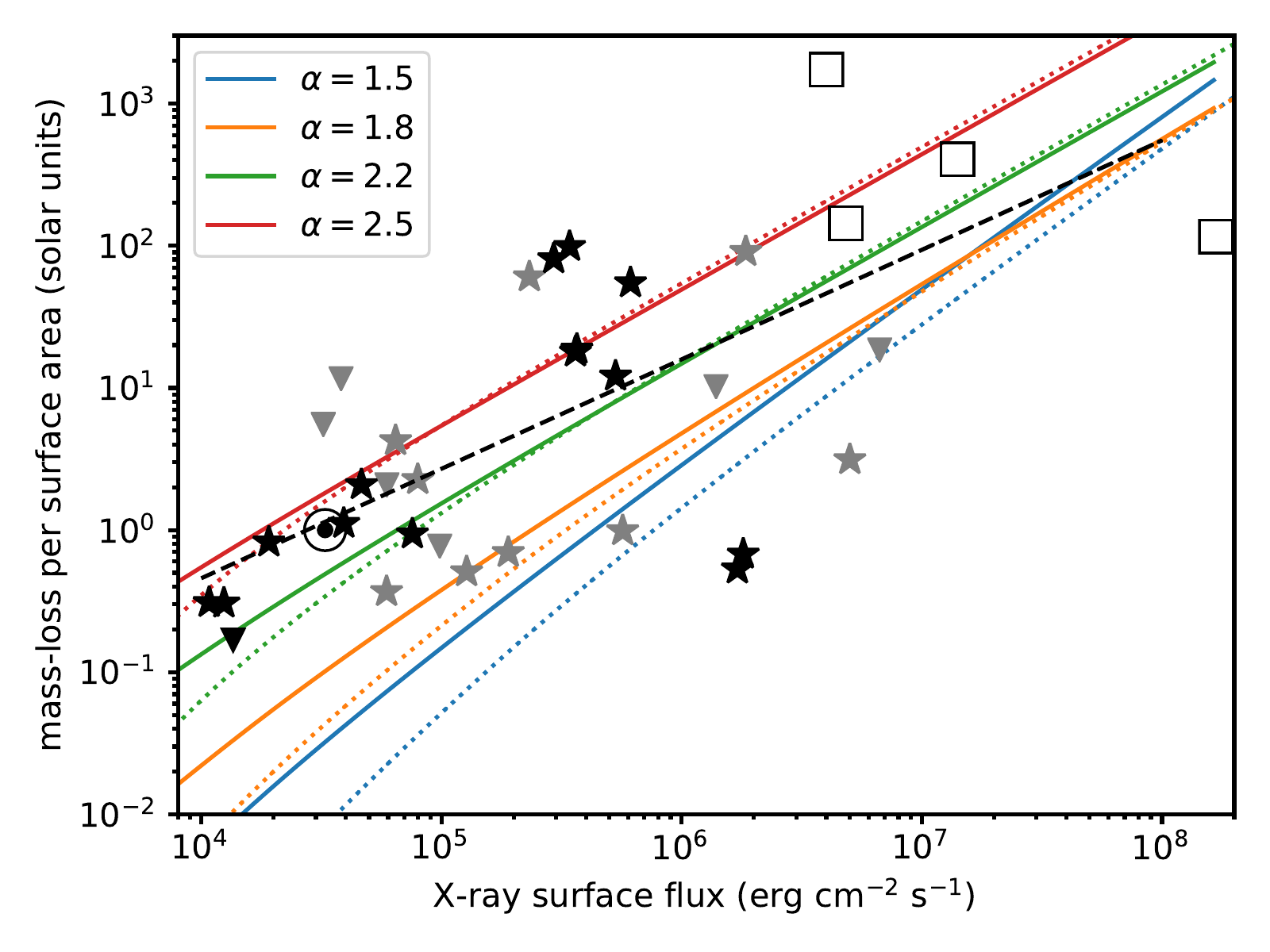}}
\vspace*{-0.1cm}
\caption{Comparison of modelled stellar CME mass-loss rates with a recent compilation of stellar mass-loss measurements. The solid and dashed lines are the expected CME mass-loss rates from the empirical model of \citet{Odert2017} for different flare power-law indices (color-coded), where solid lines are normalized to the surface area of a Sun-like star ($R_*=1R_\odot$) and dotted lines correspond to an M dwarf ($R_*=0.3R_\odot$). Star symbols correspond to the compilation of mass-loss measurements from \citet{Wood2021} for G and K dwarfs (black) and M dwarfs (gray), down-pointing triangles denote upper limits for the respective spectral type groups. Empty squares indicate estimated slingshot prominence mass-loss rates for some young fast-rotating stars \citep{Jardine2019}. The black-dashed line displays the fit between mass-loss rate and X-ray surface flux from \citet{Wood2021}.}
\label{figwood}
\end{figure}
to drop above a certain surface X-ray flux. However, recent mass-loss measurements revealed several active stars with higher mass-loss rates than previously found \citep{Wood2021} which are in better agreement with predictions from \citet{Odert2017} and could thus have a CME-dominated mass-loss (Fig.~\ref{figwood}). A slightly different approach was taken by \citet{Osten2015} who assumed energy equipartition between bolometric flare radiation and kinetic energy of the associated CME (based on solar observations) instead of extrapolating the solar flare-CME relationship. By applying their method to several stars with observed flare statistics, they found CME mass-loss rates generally comparable to previous studies. Yet another approach was developed by \citet{Cranmer2017}, who did not link CMEs with flares, but used a relationship between the surface-averaged magnetic flux with the mean kinetic energy flux of CMEs on the Sun. The mass-loss by CMEs thus scales with the magnetic 
filling factor of a star, and the model predicts that in solar-mass stars mass-loss is dominated by CMEs during the first few Gyr. \citet{Savanov2020} applied the CME mass-flare energy relation from \citet{Aarnio2012} to observed flares from Kepler to estimate typical CME masses as a function of the stellar effective temperature.\\
\begin{figure*}[htp]
\vspace*{-0.1cm}
\centerline{\includegraphics[width=\textwidth, keepaspectratio]{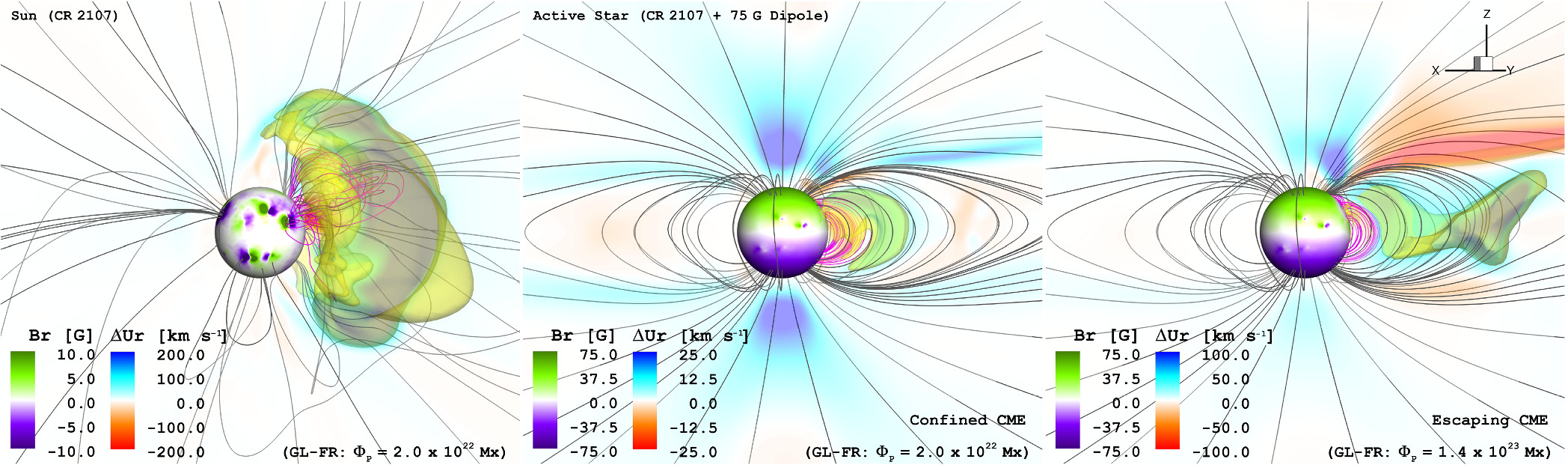}}
\vspace*{-0.1cm}
\caption{Simulations of a solar CME (left panel), a solar CME within an enhanced magnetic field (75~G, middle panel), and a more energetic CME within the enhanced field (right panel). The solar CME is confined in the stronger field, but the more energetic one can erupt. Taken from \citet{Alvarado2022b}.}
\label{figa2022}
\end{figure*}
The theoretical approaches described above predict that active stars should have a high rate of CMEs, yet observations of CMEs are still few. This discrepancy 
motivated various modeling approaches to evaluate which physical processes could be responsible that active stars could have either less CMEs than expected, or if the observational signatures may be weaker than expected and thus not detectable with current instrumentation.

\subsection{Models of observable signatures}
One potential signature to search for stellar CMEs are radio type II bursts (see Section~\ref{radiomethod}). However, non-detection of these events has led to the suggestion that shocks may not be formed by CMEs around active stars due to the large Alfv\'en speeds \citep{Mullan2019}. This was addressed in more detail by \citet{Alvarado2020} using numerical simulations, which showed that the region where shocks are formed and type II bursts can be generated are located farther away from an active star than on the Sun, resulting in lower emission frequencies which shifts them below the ionospheric cutoff, meaning 
that stellar type II bursts may only be observable from space. \citet{OFionnagain2022} predicted type II burst
occurrence over the activity cycle of the K dwarf $\epsilon$~Eri, which does not have such a strong magnetic field like a highly active star, and therefore, a potential confinement of CMEs should be less severe and the detection of associated type II bursts more promising. They found no significant cycle dependence, but stressed that the location of the CME relative to the stellar magnetic field is important, influencing the intensity and duration of the associated type II bursts.\\
Other studies aimed at modeling the optical spectroscopic signatures of stellar CMEs, thereby aiding to constrain physical parameters of CMEs for detected signatures, or inferring upper limits in case of non-detections. \citet{Odert2020} developed a simple radiative transfer model to predict Balmer line signatures of stellar CMEs. They distinguished between filament (on-disk) and prominence (off-disk) geometries and studied the dependence of the inferred signals on stellar spectral type and prominence parameters. They found that the later the spectral type the lower the required signal-to-noise ratio (S/N) for detecting a CME of a given mass due to the better contrast. Combining with predictions of the intrinsic stellar CME rates \citep{Odert2017}, it was shown that M dwarfs have the largest fraction of observable-to-intrinsic CMEs, although the required observing time necessary to detect CMEs is larger because the rate of CMEs is lower for M dwarfs at the same activity level than for solar-like stars (since the intrinsic rates scale with X-ray luminosity). The model of \citet{Odert2020} assumed that the prominence source function in the Balmer lines is dominated by scattering, like in solar prominences. However, recent results suggest that this assumption may be too simple. \citet{Leitzinger2022} adapted an NLTE radiative transfer code initially developed for the Sun \citep[e.g.][]{Heinzel1995, Heinzel1999} to Balmer line observations of a CME on an M dwarf \citep{Vida2016}. They modeled a grid of possible solutions which simultaneously reproduced the detected emission signatures in both the H$\alpha$ and H$\beta$ lines. Valid solutions were found both for prominence and filament geometry, where the latter would not be possible if the source function is due to scattering like in solar prominences, as scattering can only produce absorption signatures for filament geometry. This suggests that in M dwarfs the thermal emission of the prominence must be taken into account, which also explains why practically all CMEs (or candidate events) on M dwarfs were observed as emission signatures in the Balmer lines, as on M dwarfs erupting filaments can be seen as emission features in Balmer line spectra.\\
\citet{Wilson2022} tried to estimate the potential of UV line diagnostics for stellar CMEs, as a larger sample of spectral lines observed simultaneously could give better constraints on the plasma parameters, e.g by using line ratios for density diagnostics. Using solar CME observations from UVCS and extrapolating the line strengths to more massive CMEs as may be observed around distant stars, they identified three lines (C\,IV\,1550\AA, O\,VI\,1032\AA, C\,III\,977\AA) as the most promising. Interestingly, the signature of a potential CME in the O\,VI line was already identified in the past around the active M dwarf AD~Leo \citep{Leitzinger2011a}.\\
\citet{Cully1994} developed a model of an expanding CME to interpret the light curve of a large flare in EUV observations of AU~Mic. They could reproduce both its long tail and the spectra during the decay with a CME that has a mass of about 10$^{20}$~g. However, \citet{Katsova1999} presented an alternative interpretation involving post-eruptive energy release for this observation, without the need for a CME.\\
As coronal dimmings are a common signature of CMEs on the Sun \citep{Harra2016}, it is important to know if and how they would differ if observed on another star. \citet{Jin2020} employed a modeling framework initially developed for the Sun \citep{Jin2016} to study the ejection of a flux rope embedded within a global magnetic field to evaluate the corresponding EUV flux and emission measure evolution. They performed simulations for a variety of stellar magnetic field strengths and flux rope energies, and found that for more active stars with stronger fields (and thus higher coronal temperatures) dimmings should appear at higher temperatures as well. Dimming signatures were also simulated in some other studies discussed in the next section.
\subsection{Physical models}
One physical process which could lead to the reduction of CME rates in active stars is coronal confinement, i.e. magnetic fields overlying the flaring region could prevent the eruption of a CME. This is frequently observed on the Sun as so-called failed prominence eruptions \citep[e.g.][]{Ji2003, Shen2011}. Several studies addressed this possibility by generating simulations of the coronal magnetic field based on observed surface magnetic field maps of active stars and placing an unstable flux rope resembling a CME into the simulation. \citet{Drake2016} found that flux ropes with parameters similar to the most massive CMEs on the Sun are not able to escape from an active star like AB~Dor. Simulations of a Sun-like star with a dipolar field of 75\,G \citep{Alvarado2018} showed that most Sun-like CMEs would be confined (see Fig.~\ref{figa2022}), only much more energetic CMEs are able to escape, but the strong field leads to a reduction of their speeds. A subsequent study modelled CMEs on an M dwarf (here taken to be similar to Proxima Cen) with different assumed magnetic field strengths leading to weakly, partially, and fully suppressed events \citep{Alvarado2019}. They also determined associated dimming signatures in EUV lines, which however, appeared for all events, even the confined ones, in contrast to the majority of solar observations \citep{Veronig2021} and similar models for the Sun and Sun-like stars \citep{Jin2016, Jin2020}. Moreover, \citet{Alvarado2019} identified X-ray signatures distinguishing the different types of events which may be observable by next-generation X-ray observatories. Recently, \citet{Alvarado2022} applied their modelling framework to the active young planet-hosting M dwarf AU~Mic.\\
Whereas the studies described above addressed confinement by a global magnetic field, the local background field from an active region itself may already lead to confinement \citep[e.g.][]{Li2021}, as is the typical case for the Sun. This was considered by \citet{Sun2022} who evaluated the conditions for torus instability above stellar active regions, which is thought to be a driver for CMEs on the Sun. An expanding magnetic flux rope needs to reach a certain height above its origin where the conditions for torus instability are fulfilled to successfully accelerate and escape from the star. \citet{Sun2022} investigated the location and properties of torus-stable zones above a bipolar stellar active region which is embedded within a global dipolar magnetic field. In such an idealized scenario, two main parameter regions emerged, namely dipole- and spot-dominated regions. In the dipole-dominated regime, the global field determined the critical height where torus instability of a rising magnetic flux rope sets in, but an increasing spot size could lower the critical height. In the spot-dominated regime, the critical height increased with spot size. For small spots and moderate dipole fields, a second torus-stable zone emerged at larger heights apart from the zone immediately above the active region.\\
\citet{Lynch2019} aimed at simulating the most extreme case of a superflare and an associated CME on the young Sun-like star $\kappa^1$~Cet with a 3D magnetohydrodynamic (MHD) model. Inducing large-scale stresses to the \textit{global} stellar magnetic field via surface flows led to the gradual accumulation of free magnetic energy. Eventually, the energized field rose and reconnection set in, leading to eruption of the twisted flux rope structure and formation of a global system of post-flare loops.\\
Another relevant aspect of stellar CMEs is how they propagate away from the star. Deflection by the global magnetic field influences their trajectories and may therefore alter the CME impact rate on planets. \citet{Kay2016} adapted a solar CME deflection model to study this effect in the magnetic field of the active M dwarf V374~Peg. The large magnetic field strength led to a much stronger deflection of CMEs towards the astrospheric current sheet than on the Sun, with low-mass CMEs even being trapped there. Therefore, CME impact rates on planets would be largest if a planet's orbit is not inclined relative to the astrospheric current sheet, decreasing for increasing inclinations. A similar, but much less pronounced effect was found for impact rates on hot Jupiters orbiting Sun-like stars. For the magnetic field of the young solar analog $\kappa^1$~Cet, \citet{Kay2019} found that CMEs are efficiently deflected towards the current sheet, i.e. close to the ecliptic plane, which may result in high expected impact rates on orbiting planets.

\section{\uppercase{Stellar prominences}}
\label{stellprom}
\begin{figure}[htp]
\vspace*{-0.1cm}
\centerline{\includegraphics[width=\columnwidth, keepaspectratio]{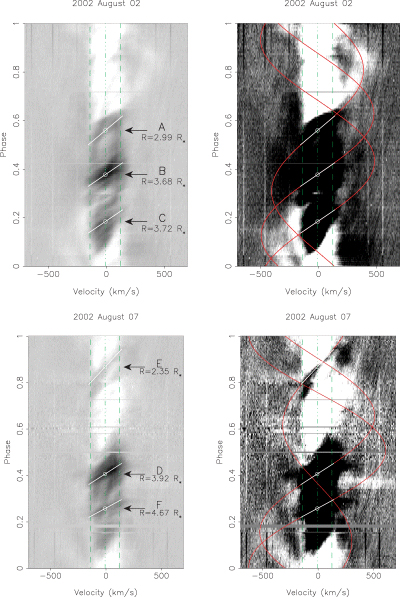}}
\vspace*{-0.1cm}
\caption{Dynamic H$\alpha$ spectra of the K dwarf BO~Mic displaying prominence signatures. The different intensity scalings highlight on-disk absorption signatures (left panels), as well as off-disk emissions (right panels) from the prominences. Taken from \citet{Dunstone2006b}, originally their Fig. 6 in \textit{The coronal structure of Speedy Mic - II. Prominence masses and off-disc emission}, Monthly Notices of the Royal Astronomical Society, vol. 373, p.
1317..}
\label{figprom}
\end{figure}

Prominences are closely connected with CMEs on the Sun, because if they become unstable and erupt they often lead to CMEs \citep[e.g.][]{Gopalswamy2003}. Therefore, the evidence for prominences on other stars may also hint at the existence of CMEs, similar to flaring activity. 
Stellar prominences have been detected as transient absorption features traveling across the rotationally-broadened profiles of chromospheric lines. In a few cases, emission features were also detected when prominences are located off-disk, similar to the Sun (Fig.~\ref{figprom}). 
The time evolution of these features shows a characteristic drifting pattern from blue to red in dynamic spectra, following the radial velocity curve of plasma clouds co-rotating with the star. The earliest detection was reported on the active young K dwarf AB~Dor where several absorption features were found in the H$\alpha$ line \citep{Robinson1986}. Subsequent studies of this star revealed that there are always several of such prominences present, with heights of several stellar radii (even beyond the Keplerian co-rotation radius) and projected areas covering up to about 20\% of the stellar disk, but temperatures and densities rather similar to solar prominences \citep{CollierCameron1989a, CollierCameron1989b}. Subsequently, the prominences were also detected in other chromospheric lines, namely Ca\,II\,H\&K and Mg\,II\,h\&k \citep{CollierCameron1990}, allowing better constraints on the prominence parameters, resulting in typical masses of a few $10^{17}$\,g. Subsequently, more fast-rotating stars with prominences were detected, such as the M dwarfs HK~Aqr \citep{Doyle1990, Byrne1996, Leitzinger2016} and EY~Dra \citep{Eibe1998}, the K dwarfs BO~Mic \citep{Jeffries1993, Dunstone2006, Dunstone2006b} and PZ~Tel \citep{Barnes2000, Leitzinger2016}, several G-type stars in the $\alpha$~Persei cluster \citep{CollierCameron1992, Cang2020, Cang2021}, as well as the post-T~Tauri star LQ~Lup \citep{Donati2000}. 
On the latter, the system of prominences does not transit the star, but is observed off-disk in emission due to the low stellar inclination. Surprisingly, no prominence-related absorption signatures have been detected on the fast-rotating K dwarf LO~Peg; however, since the star has a low inclination as well, it has been proposed that it could still have a system of non-transiting prominences, but their emission may be too faint for detection \citep{Eibe1999}. Apart from these normal (pre-)main-sequence stars, signatures of prominences 
have also been found in close binaries, such as the pre-cataclysmic binaries V471~Tau \citep{Guinan1986, Zaire2021} and QS~Vir \citep{Parsons2011, Parsons2016}.\\
Most of these prominences were found to be located close to or even outside the Keplerian co-rotation radius $R_K$. Thus, it was proposed that their stability is supported by magnetic tension of the loops in which the plasma is confined \citep{Jardine2001, Waugh2019}. These so-called ``slingshot prominences'' are not embedded in the corona like solar prominences, but above it within the stellar 
wind region \citep{Jardine2005}. They are fed by the stellar wind and if the sonic point of the flow lies below the loop summit, they will eventually reach a critical mass where confinement breaks down and the plasma will either fall back to the surface (if below $R_K$) or be 
ejected from the star \citep[if above $R_K$;][]{Jardine2019}. Typical prominence lifetimes are in the order of hours to days \citep{VillarrealDAngelo2018, VillarrealDAngelo2019}, and estimated mass-loss rates range between about three orders of magnitude below and above $\dot{M}_\odot$ \citep{Jardine2020, Waugh2021} depending on the magnetic field properties of the star. Although such slingshot prominences contribute to stellar mass-loss when erupting and may lead to impacts on planets, they do not lead to CMEs per definition, as they are already formed above the corona. However, models showed that fast-rotating stars can form both small solar-like and large slingshot prominences \citep{Waugh2022}, which was also found in some observations \citep[e.g.][]{Leitzinger2016}.

\section{\uppercase{Discussion and conclusion}}
\label{disc}
In this section we discuss the strengths and weaknesses of the different methodologies presented so far.\\
The method of Doppler-shifted emission/absorption requires spectroscopic observations as otherwise the Doppler signal can not be measured. In principle, the detectability of signatures of stellar CMEs using this approach strongly depends on the S/N of the data and the emitted or absorbed flux of the stellar CME. The spectral resolution does not play such a crucial role for this method, because the detected features are usually broad (see Section~\ref{Dopplermethod}), which can be explained by expansion during propagation, as it is the case on the Sun. However, the higher the S/N of the observations obtained within a reasonable integration time the higher the chance to detect signatures of stellar CMEs.\\
As one can see from Section~\ref{Dopplermethod}, the observing strategies are manifold. For all strategies the same rules apply. The shorter the total observing time the lower the chances to detect signatures of stellar CMEs. Single star observations are less efficient and very time consuming. In a competition-based observing time distribution (e.g. ESO, Optical Infrared Co-ordination Network/OPTICON), observing times of several nights are unlikely to be granted as such observing runs block a telescope completely. Unless one has no guaranteed-time observations at a telescope with proper instrumentation, other strategies need to be developed. One of those is the usage of multi-object spectroscopy, where a number of stars can be observed simultaneously. The draw-backs are that one needs access to a telescope with such a device and, furthermore, suitable targets which fit into the field-of-view (FoV) of the instrument. Multi-object spectrographs, such as e.g. the Fibre Large Array Multi Element Spectrograph (FLAMES) feeding the Ultraviolet and Visual Echelle Spectrograph (UVES) and GIRAFFE on the Unit Telescope 2 (UT2) of the Very Large Telescope (VLT) situated on Paranal, in the Atacama desert in Chile, or The Two Degree Field system ("2dF") installed at the Anglo-Australian-Telescope (AAT), are mounted on telescopes with a large aperture and are technically sophisticated instruments, therefore the chances of getting observing time for longer time series are relatively low. As mentioned above, when using the strategy of multi-object observations, one needs a set of targets that fit into the FoV of the instruments which restricts the target selection, in the case of the search of stellar CMEs, to young open clusters or associations/star forming regions (e.g. Chamaeleon association, Orion nebula cluster), because those stars should have the same young age and therefore a similar activity level. In young associations and star forming regions one finds young stellar objects (YSOs), classical T-Tauri stars (CTTS), but also WTTS. The latter are interesting objects to search for CMEs, because those stars are young and active and they have only a thin or already dissolved disk \citep[e.g.][]{Feigelson1999}. The latter is important for the interpretation of asymmetries in Balmer lines because such can also be caused by disk-related processes (outflows, accretion, etc.) in CTTS, which makes an interpretation very difficult.\\
Another strategy is described in \citet{Hanslmeier2017}, where the authors proposed a flare-alert system. Alert-systems in astronomy are a common approach to save observing time when aiming to investigate sporadic phenomena, such as e.g. gamma-ray bursts. This strategy is an efficient alternative to spectroscopic monitoring and relies on the association between flares and CMEs as it is known from the Sun. Especially for the investigation of stellar CMEs a flare-alert system may boost the detections to new levels. For the implementation of a flare-alert system, one needs a facility with a large FoV which is doing optical photometry and a number of available observatories hosting spectrographs which can perform spectroscopic follow-up observations. Monitoring photometrically hundreds to thousands of stars simultaneously increases significantly the probability of detecting flares. These photometric observations need to be analysed in real-time, so that when the impulsive phase (the rise) of a flare is recognised, an alert is sent to the collaborating observatories which then interrupt their ongoing observations and point immediately to the alert-target and start obtaining spectroscopic observations. So far, at least to our knowledge, such a system never went into operation as a tool to search for stellar CMEs.\\
The advantage of the method of Doppler-shifted emission/absorption is, at least in the optical domain, that observations can be carried out by the numerous ground-based observing facilities world-wide. For the detection of a prominence eruption even small to mid-sized telescopes are already sufficient for bright and active stars. Of course, as the detection mainly depends on the S/N of the data, observations with low S/N allow the detection of the most massive prominence eruptions only \citep{Odert2020}. To deduce parameters of the detected events, additional tools are necessary, except for the velocity of the 
\begin{figure}[htp]
\vspace*{-0.1cm}
\centerline{\includegraphics[width=\columnwidth, keepaspectratio]{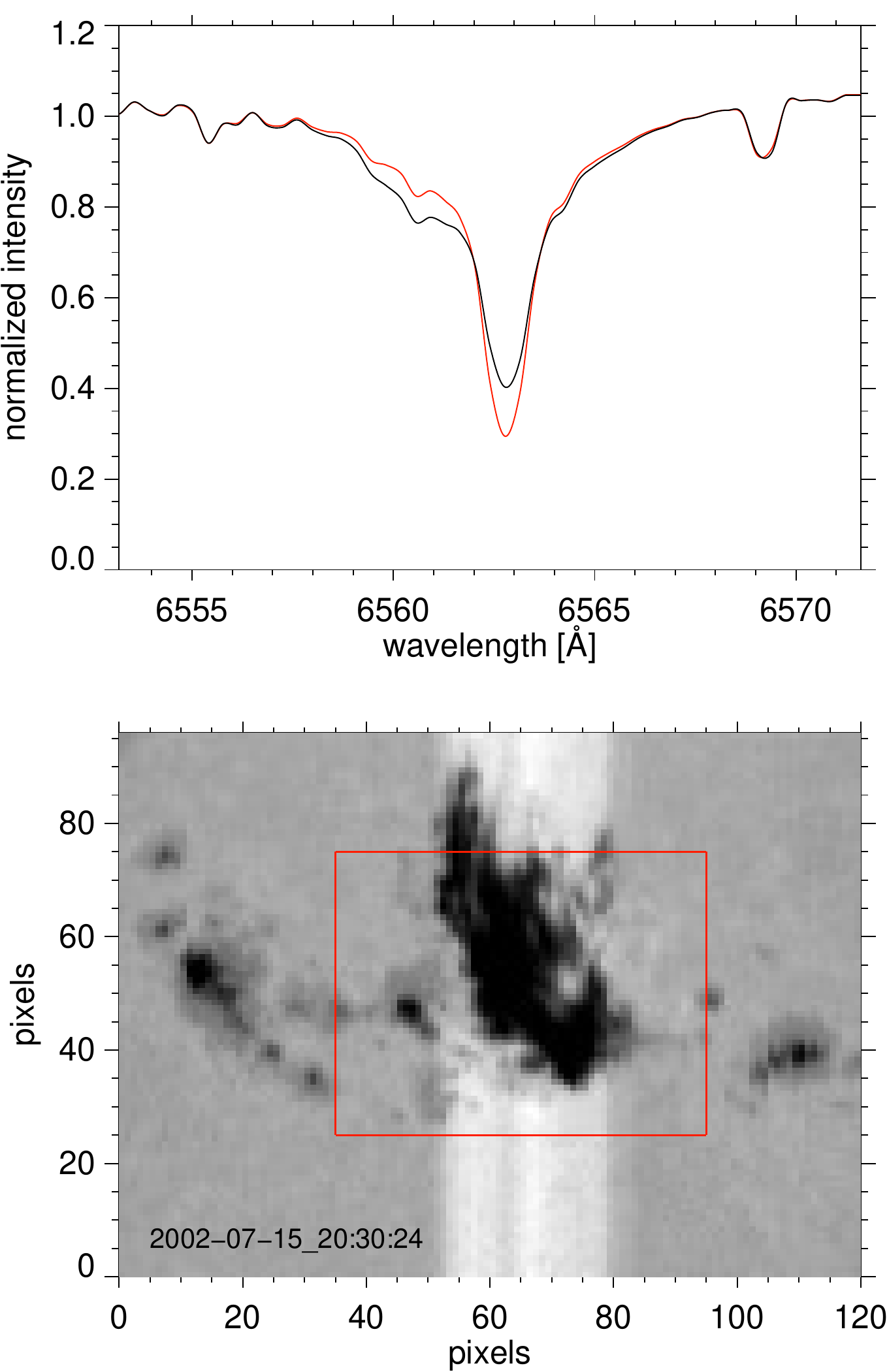}}
\vspace*{-0.1cm}
\caption{Spatially integrated solar H$\alpha$ spectra from MCCD of the quiet Sun (red) and an erupting filament (black), occurring on the Sun on 15th of July 2002 as shown in the upper panel. Corresponding 2D-image of this event observed in the blue wing of H$\alpha$ (lower panel) occurring in the decaying tail of an X-class flare after a second smaller flare. The filament spectrum was constructed from the region indicated by the red rectangle in the lower panel. Adapted from \citet{Leitzinger2021}.}
\label{figmccd}
\end{figure}
event, which can be directly measured from the data, but only in projection. This makes it generally difficult to evaluate the true nature of the event, as features at low projected velocities may be due to either rising prominences at small true velocities that do not escape or due to prominence eruptions seen in projection with larger true velocities escaping from the star. In the studies by \citet{Fuhrmeister2018} and \citet{Vida2019} numerous events have been reported with low projected 
velocities, which, therefore, can not be unambiguously identified as eruptions which indeed left the stars. Therefore, further methodologies need to be applied to determine if the signatures represent eruptions which left the star or not. One approach, which has been proposed recently \citep{Leitzinger2021, Leitzinger2022b, Namekata2022, Namekata2022b,Otsu2022, Otsu2022b} and which is known as the Sun-as-a-star concept, is the investigation of disk-integrated (Sun-as-a-star) Balmer line spectra of erupting prominences and flares. Several authors often interpreted blue-wing emissions accompanying flares as chromospheric evaporation. Up to now, it has not been proven that chromospheric evaporation is indeed visible in disk-integrated solar Balmer line spectra with typical stellar integration times of a few minutes. Moreover, there have been numerous detections of red wing enhancements during stellar flares (see Section~\ref{Dopplermethod}), where also the discrimination between eruptive prominences (ejected backwards, i.e. away from the observer) and flare-related processes, such as chromospheric condensation, is very difficult. This can be answered by a thorough analysis of Sun-as-a-star spectra of flares and eruptive prominences. Moreover, the investigation of solar observations will reveal the evolution of such eruptions in greater detail than would ever be possible for stars, especially as the temporal resolution of solar observations is much higher than for observations of other stars. The number of available instruments doing solar 2D spectroscopy with sufficient wavelength coverage (see Fig.~\ref{figmccd}) for this science case is limited. Very good examples are the MEES CCD (MCCD) at MEES solar observatory and the Solar Dynamics Doppler Imager (SDDI) at the Solar Magnetic Activity Research Telescope (SMART) at Hida observatory. MCCD is out of operation, but has a large data archive (see Fig.~\ref{figmccd} for an example of an eruptive filament), however, no full-disk coverage was available (only a part of the disk was observed). SMART/SDDI observes the full-disk Sun routinely since 2016, and covers the same wavelength range as MCCD. This instrument is not a spectrograph, but obtains the solar full-disk spectrum using several filters. \citet{Namekata2022, Namekata2022b}, \citet{Otsu2022, Otsu2022b}, and \citet{Leitzinger2021} already presented some examples of filament eruptions and flares observed in Sun-as-a-star mode.\\
To assess masses related to the Doppler signatures of stellar CMEs several authors used up to now the simple approach given in \citet{Houdebine1990}. For instance, \citet{Vida2016} estimated a mass of the event on V374~Peg in the order of 10$^{16}$~g. \citet{Leitzinger2022} re-investigated this event using an NLTE model, resulting in a mass distribution peaking at $\sim$4$\times$10$^{17}$~g, larger by more than an order of magnitude. \citet{Namekata2022} used a cloud model formalism with a fixed parameter set to determine the mass of the event on EK~Dra. \citet{Maehara2021} used two simple estimates for the mass determination of a blue-wing emission feature on YZ~CMi, which led to differences in mass of two orders of magnitudes. As one can see the mass determinations of stellar CMEs were so far order of magnitude estimations only, except for the study by \citet{Leitzinger2022} which used dedicated NLTE modeling, but also with this approach only a distribution of masses can be obtained, which gives values of 10$^{16}$-2$\times$10$^{18}$~g. Only with additional constraints it will be possible to further restrict stellar CME masses. If a stellar CME affects other chromospheric spectral lines, then those can be used to much better determine the stellar CME plasma parameters. As shown in \citet{Leitzinger2022}, using more than one spectral line can better constrain the plasma parameters compared to the single line observations from most of the studies.\\
Furthermore, with optical observations one underestimates the mass of the CME, as one observes the dense and cool filament, i.e. the CME core, only, as the chromospheric lines do not probe the piled-up hot coronal material. On the Sun we know that there are CMEs without embedded filaments, those would remain invisible on stars in optical spectra. A further advantage is that several (depending on the instrument) targets can be observed simultaneously and there already exists a significant number of candidate events in the literature which can further be constrained, using new methodologies (see above), to enable statistics of stellar CMEs.\\
The signature of coronal dimmings requires X-ray or EUV light curves to be identified. To further investigate the origin of these dimmings, additional spectroscopy is necessary, as shown in \citet{Veronig2021}, to evaluate plasma parameters, such as the emission measure and temperature. Of course, in comparison with the method of Doppler-shifted emission/absorption in the optical domain, the dimming signature requires short-wavelength observations which are only accessible from space and are, therefore, more difficult to obtain, especially as space observatories or satellites are much less numerous than ground-based facilities. However, the data pool of available X-ray observations is not yet fully explored.\\
The signature of coronal dimmings has revealed so far the largest number of events in one study. This sounds very promising, especially as there are still other X-ray data which can be analysed. As this signature requires X-ray imaging, several targets in the FoV can be observed at the same time, enhancing the efficiency of the approach. If the signature is detected then a CME has likely happened, as it indicates missing coronal material. The conclusions can further be strengthened by simultaneous X-ray/EUV spectroscopy. Therefore, the signature of coronal dimmings indicates more reliably a CME, as in cases with low projected velocities the Doppler method does not unambiguously indicate that plasma was ejected from the star. However, for events with large projected velocities, the Doppler method also reliably indicates stellar CMEs. Moreover, coronal dimmings have been proven to be strong also in Sun-as-a-star observations, making it a reliable indicator. The detection of coronal dimmings strongly depends on the definition of a pre-flare level, as time series have limited durations it may happen that for the definition of a pre-flare level only few data points are available. Moreover, it may also happen that the post-flare phase is not fully covered, hampering the dimming identification. Furthermore, the determination of CME parameters requires other tools, such as solar scalings and/or modelling. Finally, coronal dimmings may not be detectable in very strong flares, as the dimming signature may be outshined by the flare. In addition, in frequently flaring stars dimming signatures following one flare may be fragmented by subsequent flares.\\
The signature of radio bursts requires dynamic spectra in the radio domain. The correlation on the Sun between type II bursts and CMEs was found at decametric/metric wavelengths. These wavelength domains are covered by a limited number of telescopes, but also very large ones, such as LOFAR, the UTR-2, the Giant Meterwave Radio Telescope (GMRT), the JVLA, or ASKAP. At low frequencies the spatial resolution of the radio telescope as well as its sensitivity is limited; moving to higher frequencies increases both. As mentioned in Section~\ref{modelling}, theoretical studies revealed that on magnetically active stars type II bursts may appear only at frequencies below the ionospheric cut-off which would explain the non-detections as most campaigns focused, of course, on the most active targets. A probably more promising strategy would be to focus on moderately active stars with this method, which certainly causes more observational efforts but corresponds better to the theoretical predictions \citep{OFionnagain2022}.\\
Continuous X-ray absorption requires X-ray observations, therefore, the same data limitations apply as for the signature of coronal dimmings. Continuous X-ray absorption was not identified as a solar signature, but is a reasonable explanation of the observed column density evolution during flares by obscuration from an expanding plasma, such as a CME. The detection of continuous X-ray absorption is limited to longer lasting flares as a decay in hydrogen column density must be evident for this approach to be applied. A handful of events has been presented in the literature and there are still more X-ray data to be analysed, similarly as for coronal dimmings. In the X-ray domain one may therefore find stellar CMEs by three methods at the same time (coronal dimmings, continuous X-ray absorption, and Doppler-shifts). Detection of all signatures in one event would yield a much stronger constraint on the detection itself than by using only one method.\\ 
In the special case of the pre-cataclysmic binary V471~Tau, \citet{Bond2001} predicted CME rates of 100-500~day$^{-1}$, derived from two sequences of Si\,III absorptions. Only recently, \citet{Kovari2021} investigated the V471~Tau system, with respect to the origin of the magnetic activity of the K-dwarf component of the system. The authors used optical spectroscopic observations of V471~Tau covering nearly 29~h to perform Doppler imaging, and thereby locating spots on the K-dwarf component. The dynamic spectra shown in \citet{Kovari2021} revealed only surface variations, but no eruptions from the system. If the system indeed had a CME rate in the range of 100-500 CMEs per day, then a few events would have been expected in the optical spectroscopic observations.\\
As discussed above, every method presently used to search for stellar CMEs has, of course, its advantages and drawbacks. One step forward would be the approach of multi-wavelength campaigns combining the different signatures to identify stellar CMEs unambiguously and to better determine their parameters. However, this requires a significant observational effort and has usually a low probability to succeed. The organisation of a multi-wavelength campaign requires different ground- and space-based observatories which must operate coordinated for at least several hours to hopefully catch a transient phenomenon like a CME. Especially for flare star investigations such campaigns have been already successfully carried out \citep[e.g.][]{Hawley1991,Lalitha2020, Paudel2021}. Although such campaigns will probably not enhance stellar CME statistics significantly, they will certainly contribute to a deeper understanding of stellar CMEs and their observable signatures.\\
As one can see from the previous sections, there have been many studies using different detection methods to reveal the occurrence rate and parameters of stellar CMEs. For more than 30 years now stellar CMEs are a subject of active research, especially in the last decade the number of publications on this topic has significantly increased. However despite the efforts during the past 30 years still we do not have a quarter as good statistics on stellar CME parameters as on the Sun. Because of the implications of CMEs for planetary habitability and stellar evolution, both in our solar system as well as predicted for extrasolar systems, obtaining reliable stellar CME parameter statistics is a very important goal.\\
There is ongoing research using the established methodology, and recently new methodologies have been (coronal dimmings) and will be (Sun-as-a-star optical spectroscopy) established. Numerous theoretical approaches and modeling frameworks (incl. MHD, NLTE) have been successfully applied to the stellar CME case and highly interesting results were found explaining observational non-detections, refining observational signatures, as well as improving the CME parameter determination. All in all, there will be many more interesting publications on this topic in the future which will certainly help to improve our knowledge on this sporadic phenomenon with its important implications.


\acknowledgements{M.L. and P.O. acknowledge the Austrian Science Fund
(FWF): P30949-N36, I5711-N for supporting this research. We thank all the authors for their consent on reusing their published figures in this review.}

\newcommand\eprint{in press }

\bibsep=0pt

\bibliographystyle{aa_url_saj}

{\small

\bibliography{Mybibfile.bib}
}

\begin{strip}

\end{strip}

\clearpage

{\ }

\end{document}